\documentclass[twocolumn]{aastex7}
\usepackage{amsmath}
\usepackage{CJK}
\usepackage{booktabs}
\usepackage{tabularx}

\defcitealias{bgagliuffi2025jwst}{BGB25}
\newcommand{\cta}{\citetalias}

\begin{document}
\begin{CJK*}{UTF8}{gbsn}
\title{The Architecture of the 14 Herculis System Suggests Primordial Ejection of a Massive Planet}

\author[0000-0003-0834-8645]{Tiger Lu (陆均)}
\altaffiliation{Flatiron Research Fellow}
\affiliation{Center for Computational Astrophysics, Flatiron Institute, 162 5th Avenue, New York, NY 10010, USA}
\affiliation{Department of Astronomy, Yale University, New Haven, CT 06511, USA}
\email[show]{tlu@flatironinstitute.org}  

\author[0000-0003-3130-2282]{Sarah C. Millholland}
\affil{Department of Physics, Massachusetts Institute of Technology, Cambridge, MA 02139, USA}
\affiliation{MIT Kavli Institute for Astrophysics and Space Research, Massachusetts Institute of Technology, Cambridge, MA 02139, USA}
\email{sarah.millholland@mit.edu}

\author[0000-0002-7670-670X]{Malena Rice}
\affiliation{Department of Astronomy, Yale University, New Haven, CT 06511, USA}
\email{malena.rice@yale.edu}

\author[orcid=0000-0003-2338-2091]{Brennen Black}
\affiliation{Tsung-Dao Lee Institute, Shanghai Jiao-Tong University, Shanghai, 520 Shengrong Road, 201210, China}
\email{brennenblack@sjtu.edu.cn}

\author[0000-0001-8170-7072]{Daniella C. Bardalez Gagliuffi}
\affiliation{Department of Physics \& Astronomy, Amherst College, 25 East Drive, Amherst, MA 01002, USA}
\affiliation{American Museum of Natural History, Department of Astrophysics, 200 Central Park West, New York, NY 10024, USA}
\email{dbardalezgagliuffi@amherst.edu}

\author[0000-0001-6396-8439]{William O. Balmer}
\affiliation{Department of Physics \& Astronomy, Johns Hopkins University, Baltimore, MD 21218, USA}
\affiliation{Space Telescope Science Institute, 3700 San Martin Dr., Baltimore, MD 21218, USA}
\email{wbalmer@stsci.edu}

\author[0000-0003-3818-408X]{Laurent Pueyo}
\affiliation{Space Telescope Science Institute, 3700 San Martin Dr., Baltimore, MD 21218, USA}
\email{pueyo@stsci.edu}

\author[0000-0002-0078-5288]{Mark R. Giovinazzi}
\affiliation{Department of Physics \& Astronomy, Amherst College, 25 East Drive, Amherst, MA 01002, USA}
\email{mgiovinazzi@amherst.edu}

\author[0000-0003-2630-8073]{Timothy D. Brandt}
\affiliation{Space Telescope Science Institute, 3700 San Martin Dr., Baltimore, MD 21218, USA}
\email{tbrandt@stsci.edu}

\begin{abstract}

The 14 Herculis system hosts two super-Jupiters on eccentric, significantly misaligned orbits. This orbital architecture represents a dynamical puzzle that demands explanation. In this work, we reproduce the system's dynamical history and current architecture using a large suite of \textit{N}-body simulations of planet-planet scattering. Our results demonstrate that planet-planet scattering is able to reproduce 14 Her's peculiar orbital architecture, but only if additional massive bodies were initially present in the system that were subsequently ejected. The mass of any such ejected planet can in turn constrain the system's initial configuration. We also analyze the present-day secular evolution of the system and conclude that while there are most likely nontrivial eccentricity and inclination oscillations currently occurring, the magnitudes of these oscillations are not strong enough to allow tidal forces to meaningfully alter the system's architecture. Finally, we discuss how forthcoming observations from future \textit{Gaia} data releases and the \textit{Roman} mission may situate 14 Her's dynamical history within a broader, population-level framework.

\end{abstract}

\section{Introduction}
The planets of our own Solar System reside in nearly circular and coplanar orbits, the expected outcome of formation within a dissipative protoplanetary disk\footnote{Though this configuration does not precude a dynamically hot history, e.g. \citet{nesvorny2011young}.} \citep{goldreich1980disk, cresswell2007evolution, armitage2010pf}. Over the past decade, missions such as \textit{Kepler} \citep{borucki_2010} and \textit{TESS} \citep{ricker2015tess} have uncovered a wide range of exoplanetary system architectures that do not fit neatly into this paradigm. Our view of planetary system formation has expanded, and it is now believed that subsequent dynamical instabilities are responsible for generating many of the fascinating orbital architectures we observe today.

Due to geometric and sensitivity biases inherent in transit and radial velocity surveys, our knowledge of exoplanet demographics is most complete for close-in and massive planets. We therefore have a reasonably good, though incomplete, understanding of the processes that sculpt close-in planetary systems, and dynamical instabilities are capable of reproducing many aspects of their observed demographics \citep[e.g.][]{juric_tremaine_2008, Pu15, izidoro2021chains}. The same cannot be said for their more widely-separated counterparts; basic aspects of the demographics of distant multi-planet systems remain poorly understood. 

There is evidence to suggest that the dynamics may be qualitatively different in these two regimes. For instance, giant planets are expected to preferentially form beyond the water-ice line \citep{lin1996orbital, ida2004deterministic}. Thus we expect more massive planets in outer regions of planetary systems, peaking at the water-ice line and dropping off with orbital separation \citep{fernandes2019turnover, wittenmyer2020cool, fulton2021cls}. Dynamical interactions between close-in planets are also expected to be collision-dominated, while similar violent interactions further out in a planetary system will preferentially lead to ejections \citep{safronov1972evolution}.

While the data presently does not support population-level insights into the outer reaches of planetary systems, intriguing individual systems have been characterized. Of particular relevance to this work are multi-planet systems which host eccentric and mutually misaligned planets such as HAT-P-11 \citep{An2025significant}, $\pi$ Men \citep{XuanWyatt2020} and HD 73344 \citep{ZhangWeiss2025}. These characteristics are typically viewed as a smoking gun for dynamically turbulent pasts.

The 14 Herculis (hereafter 14 Her) system adds to this census. 14 Her is a near-solar-mass K0V star \citep{perkins1989catalog} 17.9 pc away \citep{gaiadr3} hosting two super-Jupiters. A roughly five-year recurring RV signature was attributed to the inner planet by \cite{butler2003keck}. A long-term trend \citep{wittenmyer2007long} was later confirmed as a widely-separated outer planet by \cite{bgagliuffi2021her}, whose properties were refined with the addition of astrometric data from the Hubble Space Telescope's Fine Guidance Sensor by \cite{benedict2023her}. Recently, the outer planet was directly imaged with JWST/NIRCam by \citealt{bgagliuffi2025jwst} (hereafter BGB25). Through a joint analysis of their JWST relative astrometry, absolute astrometry from the Hipparcos-Gaia Catalog of Accelerations \citep{brandt2018hgca}, and over 25 years of radial velocities, their study represents our most up-to-date constraints on 14 Her's orbital architecture and indicates that the two super-Jupiters' orbits are both eccentric and significantly misaligned. Due to a degeneracy in the measurement of 14 Her b's inclination, \cta{bgagliuffi2025jwst} identified two inclination modes for planet b, corresponding to two mutual inclination solutions. We refer to these as the ``prograde" ($<90^\circ$) and ``retrograde" ($>90^\circ$) family of solutions. In addition, they also provide the most up-to-date system age estimate of $4.6_{-1.3}^{+3.8}$ Gyr. Relevant dynamical quantities from their fits are reported in Table \ref{tab:params}.

\begin{deluxetable}{lccr}
\tablewidth{0pt}
\tablecaption{Relevant dynamical parameters of the 14 Her System derived by \cta{bgagliuffi2025jwst}\label{tab:params}}
\tablehead{
Parameter & Description & Units &
Value}
\startdata
\hline
      $M_\mathrm{*}$ & Mass of star & $\rm{M}_\odot$ & $0.97^{+0.04}_{+0.04}$\\
      $m_\mathrm{b}$ & Mass of b & $\rm{M}_\mathrm{J}$ & $8.9_{-1.7}^{+1.3}$\\
      $m_\mathrm{c}$ & Mass of c & $\rm{M}_\mathrm{J}$ & $7.9_{-1.2}^{+1.6}$\\
      $a_\mathrm{b}$ & Semimajor Axis & au & $2.839_{-0.041}^{+0.039}$ \\
      $e_\mathrm{b}$ & Eccentricity & $\dots$ & $0.3683_{-0.0029}^{+0.0029}$
      \\
      $a_\mathrm{c}$ & Semimajor Axis & au & $20.0_{-4.9}^{+12.0}$ \\
      $e_\mathrm{c}$ & Eccentricity & $\dots$ & $0.52_{-0.12}^{+0.16}$
      \\
      $\Theta$\tablenotemark{a} & Mutual Inclination & deg & $32.09^{+13.75}_{-14.90}$\\
      $\Theta$ & Mutual Inclination & deg & $145.0^{+10.89}_{-16.04}$\\ 
      \hline
\enddata
\tablenotetext{a}{\cta{bgagliuffi2025jwst} identified two inclination modes for planet b, corresponding to the two mutual inclination solutions. We note that \cite{xiao202514her} included \textit{Gaia} DR2 data in a new analysis and found just one mutual inclination peak at $\Theta = 35.3_{-7.3}^{+6.8}$ degrees. Upcoming data from \textit{Gaia} DR4 will be decisive in resolving this degeneracy.}
\end{deluxetable}
We note that both planets are distant enough such that tidal forces are not expected to have significantly sculpted the system in the intervening years since formation, which we explore in detail in \S \ref{sec:present}. Hence, 14 Her represents a pristine relic of the dynamical mechanism that generated its peculiar architecture. This architecture is not the expected outcome of planet formation, and demands an explanation. Together with the elevated eccentricities of the planets' orbits, the relatively high mutual inclination solutions suggest a dynamically hot history for the system.

In this work, we extensively analyze the dynamical history of the 14 Her system. We show that the system's architecture is consistent with planet-planet scattering as posited by \cta{bgagliuffi2025jwst}, but only if another giant planet was originally present in the system that has since been ejected. The paper is structured as follows. In \S \ref{sec:stab} we analyze the long-term stability of the posteriors derived by \cta{bgagliuffi2025jwst} and demonstrate that the system is unlikely to be experiencing von Zeipel-Lidov-Kozai oscillations today. In \S \ref{sec:pp_scattering} we report results from \textit{N}-body simulations tracing the dynamical history of the 14 Her system. We show that initial conditions consisting of only the two original planets in the system cannot generate the present-day architecture, but systems with three massive planets can. In \S \ref{sec:present} we analyze the present-day secular evolution of the system and show that the system is likely presently undergoing significant eccentricity and inclination oscillations. In \S \ref{sec:disc} we discuss some implications of our results for the system's formation, as well as prospects for future population-level insights with upcoming results from \textit{Gaia} and \textit{Roman}. We conclude in \S \ref{sec:conclusions}.

\section{System Stability}
\label{sec:stab}
Not all of the system parameters encompassed by the posteriors derived in \cta{bgagliuffi2025jwst} correspond to physically stable architectures. Here we assess the dynamical stability of these posteriors, and discard those which are not stable long-term.

While analytic criteria exist for the long-term stability of two coplanar massive eccentric planets \citep{hadden_lithwick_2018} and for mutually inclined systems where the inner body is treated as a test particle \citep{bhaskar2024dynamical}, the case of two massive, eccentric, mutually inclined bodies remains without such a criterion. We thus explored the stability of 14 Her via \textit{N}-body simulations. We randomly drew 10000 system configurations from the posteriors presented in \cta{bgagliuffi2025jwst}, and integrated each forward in time for $1$ Gyr. We used the \texttt{WHFast} integrator \citep{rein_2015} in \texttt{REBOUND} \citep{Rein_2012}. We adopted a timestep of $1/20$th the effective period at pericenter $\tau_\mathrm{eff}$ of 14 Her b

\begin{equation}
    \tau_\mathrm{eff} = 2 \pi \sqrt{\frac{(1-e_\mathrm{b})^3}{1+e_\mathrm{b}} \frac{a_\mathrm{b}^3}{G M_*}}, 
\end{equation}
where $m_\mathrm{p}$, $M_*$ are the masses of the planet and star, respectively, $a_\mathrm{b}$ is the planet's semimajor axis, and $e_\mathrm{b}$ its eccentricity. Such short timesteps have been shown to adequately resolve highly eccentric orbits \citep{wisdom_2015, hernandez_2022}.

There are two pathways to instability in exoplanetary systems \citep[e.g.][]{deck2012rapid, tremaine_bible}. The first is Hill instability, which is triggered by close encounters between planets. We checked for this condition via overlapping Hill radii \citep{hill1878lunar}:

\begin{equation}
    R_\mathrm{Hill} \equiv a(1-e)\left(\frac{m_\mathrm{p}}{3M_*}\right)^{1/3}.
\end{equation}
This is a common indicator for a system's immediate subsequent instability \citep[e.g.][]{obertas_2017, Tamayo20, lammers_winn_2024,lu_hip}. The second is Lagrange instability, in which secular perturbations ultimately result in collisions with the central star (which is endowed with $R_* = R_\odot$, for collision detection purposes) or the planet becoming unbound. Within the 1 Gyr integration time, $19.7\%$ of our simulations are unstable. When either Hill or Lagrange instability was detected, the simulation was halted. These systems tended to be unstable on relatively short timescales. Of the unstable simulations, $38\%$ went unstable within $10^7$ years and $63\%$ within $10^8$ years.

The remaining 8028 simulations were stable over the 1 Gyr integration time. $3\%$ of our simulations experience significant semimajor axis oscillations of $|\Delta a / a| \geq 1\%$. While this threshold is suggested as an instability metric by \cite{bhaskar2024dynamical}, each of these simulations did not exhibit major instability over the 1 Gyr integration window, so we did not rule out these configurations in the remainder of our analysis. There was no obvious pattern for the parameter space in which these relatively large semimajor axis deviations occur, but they all belonged to the prograde family of solutions. This result, as well as the genesis of the major instabilities arising from ZLK oscillations, is qualitatively in good agreement with the results of \cite{bhaskar2024dynamical}.

\begin{figure*}
    \centering
    \includegraphics{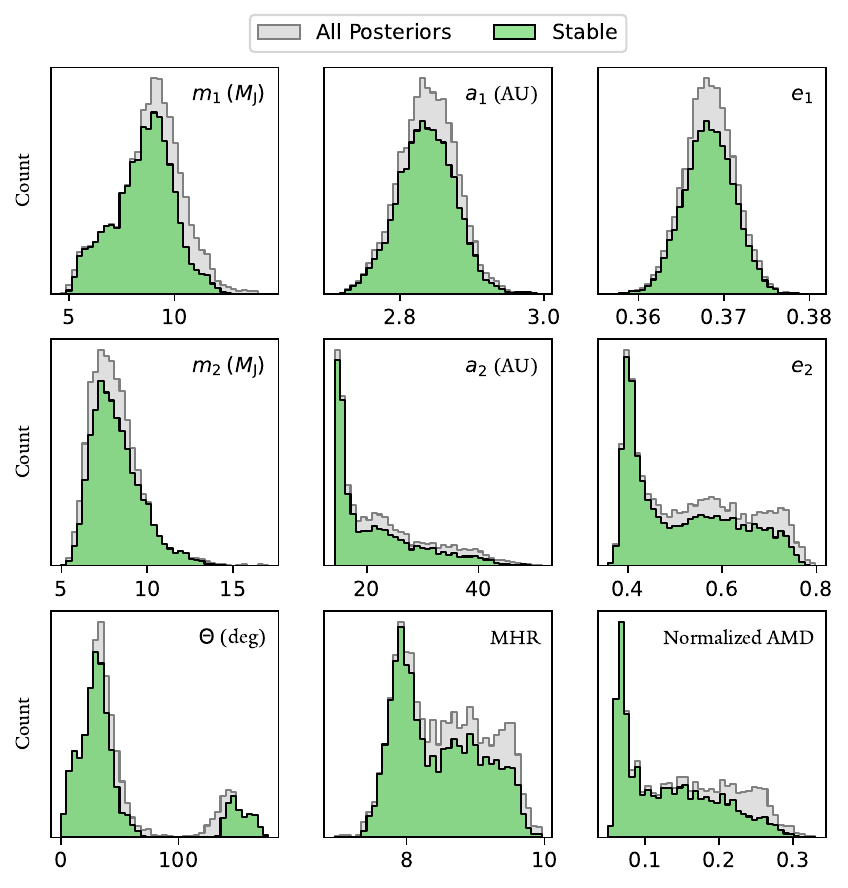}
    \caption{The posterior distributions (gray) and the subset of those which are stable (green) for nine parameters. We plot the mass, semimajor axis, and eccentricity of each planet, as well as the mutual inclination, mutual Hill radius and normalized angular momentum deficit (see Equations \eqref{eq:mhr} and \eqref{eq:namd}, respectively). Approximately $80\%$ of the posteriors are stable over 1 Gyr.}
    \label{fig:posteriors}
\end{figure*}
Figure \ref{fig:posteriors} shows the full set of original posteriors (gray), overplotted with the stable subset (green). Most of the parameter space was stable, and thus we could not definitively rule out most of the posteriors via our stability analysis. We do note a slight preference for a less massive 14 Her b, a less eccentric 14 Her c, and a smaller orbital separation between the planets. The strongest conclusions can be drawn about the mutual inclination distributions. Our results strongly disfavor near-perpendicular orbits; no configurations with $75^\circ < \Theta < 105^\circ$ are stable. 52\% of simulations with $39.2^\circ < \Theta < 140.7^\circ$ -- the nominal ``Kozai angles" between which traditional ZLK oscillations \citep{von_zeipel_1910, lidov1962evolution, kozai1962secular, naoz2016eccentric} are expected to generate significant eccentricity oscillations in one or both planets -- are unstable. This contrasts with systems with $\Theta$ outside of this range, of which $92\%$ are stable. 

In summary, stability analysis of the 14 Her system rules out near-perpendicular configurations which would exhibit extremely high-amplitude ZLK cycles. Presently ongoing ZLK cycles are disfavored, but cannot be completely ruled out. However, we emphasize that falling outside the nominal ZLK angles does not rule out significant secular eccentricity oscillations, which will be discussed in more detail in \S \ref{sec:present}.

\section{Formation via Planet-Planet Scattering}
\label{sec:pp_scattering}
There are numerous dynamical mechanisms readily capable of exciting significant eccentricities and inclinations. These include planet-planet scattering, ZLK oscillations, and secular chaos \citep{lithwick2014secular}. In this work, we focus on planet-planet scattering as a pathway for generating 14 Her's orbital architecture. Planet-planet scattering, violent impulsive gravitational interactions arising from close encounters between planets, is believed to be an ubiquitous process in the primordial stages of planetary system formation, and it has been invoked as a natural mechanism to explain the eccentricities and mutual inclinations of many exoplanetary systems \citep[e.g.][]{weidenschilling1996scattering, ford2008, juric_tremaine_2008, ford2008, chatterjee_2008, nagasawa2011orbital, beauge2012multiple, petrovich_2014, anderson_2020,marzari2025planet,lu_hatp11, dong2026scattering, esposito2026scattering}.

We have run several scattering ensembles, with the aim of reproducing potential dynamical histories of 14 Her. We describe aspects of the setup shared across all simulations here, and details of each specific simulation suite are detailed in the following sections. All simulations were performed with the hybrid integrator \texttt{TRACE} \citep{Lu_TRACE} in \texttt{REBOUND}, with a timestep set to $1/20$th the initial orbital period of the innermost planet. We tracked both ejections and collisions, and we removed any planet that exceeds a distance of $10^5$ au from the host star using the \texttt{exit\_max\_distance} functionality in \texttt{REBOUND}. We checked for collisions by flagging overlapping planetary radii each timestep and assumed perfect mergers that conserve mass and angular momentum but not energy. We scaled out the energy lost both from collisions and ejected bodies with the \texttt{track\_energy\_offset} feature.

Planet-planet scattering is a highly chaotic process, and the observational constraints on planet b are particularly stringent. Reproducing the exact present-day configuration of 14 Her would therefore require an enormous computational investment with limited additional physical insight. Instead, we focused on evaluating whether our simulated systems produce qualitatively similar architectures. 

To quantify similarity, we considered three metrics: the mutual inclination $\Theta$, the planets’ mutual separation expressed in units of the mutual Hill radius \citep[MHR;][]{chambers1998making}:

\begin{equation}
    \label{eq:mhr}
    \text{MHR} = \frac{a_{\mathrm{c}} + a_{\mathrm{b}}}{2} \left(\frac{m_{\mathrm{c}} + m_{\mathrm{b}}}{3 M_*}\right)^{1/3},
\end{equation}
and the normalized angular momentum deficit \citep{laskar1997amd, laskar2000spacing,turrini2020namd}:

\begin{equation}
    \label{eq:namd}
    \text{NAMD} = \frac{\sum_{i\in \{b,c\}}m_i\sqrt{G M_* a_i} \left(1 - \sqrt{1-e_i^2}\cos I_i\right)}{\sum_{i\in \{b,c\}}m_i\sqrt{G M_* a_i}}
\end{equation}
where $I$ denotes the planet's orbital inclination with respect to the system's invariable plane. Because the posterior distribution for planet b's inclination is bimodal, we modeled the joint distribution of $\Theta$, $e_\mathrm{12}$, and MHR using a two-component Gaussian Mixture Model (GMM), corresponding to the prograde and retrograde solution families. We computed the Mahalanobis distance $D_\mathrm{M}$ \citep{mahalanobis1936note, anderson1958introduction}, a measure of the distance between a point and a distribution, to both peaks for each simulation's state vector $\vec{x} = [\Theta, \mathrm{MHR}, \mathrm{NAMD}]$:

\begin{equation}
    D_\mathrm{M} \equiv \sqrt{(\vec{x} - \vec{\mu})^T \Sigma^{-1}(\vec{x} - \vec{\mu})}
\end{equation}
where $\vec{\mu}$ is the mean vector and $\Sigma^{-1}$ is the inverse covariance matrix inferred from the GMM fit. We report the minimum $D_\mathrm{M}$ to either peak.

\subsection{Two-Planet Scattering}
\label{sec:twobody}
The simplest scattering scenario involves an instability between 14 Her b and 14 Her c alone, with both planets surviving the scattering process. This picture does not require invoking any additional unobserved bodies and corresponds to the minimal parameter space for exploring planet-planet scattering. 

To inform our initial conditions, stability provides one powerful constraint. Assuming initially near-circular and aligned orbits, \cite{Gladman93} analytically showed that two-planet systems are Hill stable for arbitrary long timescales if their semimajor axes satisfy

\begin{equation}
    \label{eq:twop_stability}
    a_\mathrm{c} > \left[ 1 + \delta \left(\frac{m_\mathrm{c} + m_\mathrm{b}}{M_*}\right)^{1/3} \right] a_\mathrm{b},
\end{equation}
where $\delta = 2.4$ is the derived critical threshold necessary to generate instabilities. Conservation of orbital energy provides another powerful constraint: in the absence of collisions or ejections, the orbital energy of the initial configuration must match that of the present-day system. Accounting for both these constraints and using the best-fit values for the 14 Her system, we have $a_\mathrm{b,init} = 3.87$ au and $a_\mathrm{c,init} = 6.23$ au, which places the primordial 14 Her b and 14 Her c just wide of 2:1 mean motion resonance.

We ran 8028 two-planet scattering simulations, with one run associated with each of the stable posteriors derived in \S \ref{sec:stab}. Both planets were initialized on near-circular, aligned orbits, with eccentricities and inclinations drawn from $\mathcal{U}(0.01, 0.05)$ and $\mathcal{U}(0^\circ, 2^\circ)$, respectively. All other orbital angles were completely randomized. The initial semimajor axes were informed by the Hill instability and energy matching arguments explained above, where we set $\delta = 2.3$. For collision detection purposes we set $R_\mathrm{b} = R_\mathrm{c} = R_\mathrm{J}$ for each planet \citep[e.g.][]{chen2017radius}. Simulations were run for $10^8$ years, but if a collision or ejection occurred, the simulation was immediately halted. The mean energy error was $|\Delta E/E| \sim 0.08$.

\begin{figure}
    \centering
    \includegraphics{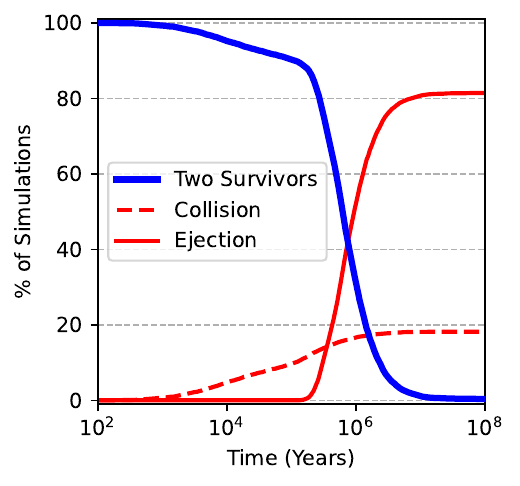}
    \caption{Branching ratios as a function of simulation time for our two-body scattering simulations. Blue, red solid, and red dashed lines represent simulations with two survivors, an ejection, and a collision, respectively. By $10^8$ years, the relative fractions of each ratio have stabilized, demonstrating that our simulations have run long enough to capture population-level dynamics. Systems with two surviving planets are highly disfavored.}
    \label{fig:br}
\end{figure}
The results of our two-planet scattering simulations are shown in Figure \ref{fig:br} as a function of integration time. At $10^8$ years the relative fraction of each simulation outcome has stabilized, an indication that our simulations have been run long enough to represent accurate population-level trends \citep{Ford_2001}. The vast majority of our simulations resulted in either an ejection $(81.4\%)$ or collision $(18.2\%)$, with only $0.4\%$ retaining both planets. Of simulations that retain both planets, none match the stable posteriors at the $1 \sigma$ level, and only one matches at the $2 \sigma$ level. 

While we cannot completely rule out two-planet scattering, our results strongly disfavor it as a formation pathway for 14 Her. It is not feasible to comprehensively explore the parameter space of viable initial conditions to completely rule out the two-planet scattering scenario, but the stability and energy conservation constraints are quite strong and significantly restrict this parameter space. For instance, the planets cannot initially be much further apart, or they would be stable over arbitrarily long timescales. Hence, we do not expect there to be other reasonable initial conditions which would lead to qualitatively different results.

\subsection{Three-Planet Scattering}
\label{sec:3p_scattering}
It is possible that other massive planets initially existed in the 14 Her system and were either ejected from the system or were thrown onto undetectable distant orbits. Indeed, it has been shown that at times it is necessary to invoke such an additional body in order to reproduce mutually inclined system architectures \citep[e.g.][]{mardling_2010, lu_hatp11}. 

We ran six simulation suites modeling planet-planet scattering between three initial planets. The setup of these simulations is as follows. Each suite consisted of $8028$ simulations, corresponding to the number of Hill stable priors from \S \ref{sec:stab}. In each simulation, two of the planets had masses drawn from the stable posteriors. The mass of the final planet was the free parameter we vary in each suite, and is reported in Table \ref{tab:threebody}. The additional planet was also endowed with $R = R_\mathrm{J}$. The order of the three planets was randomized, and the planets were initialized on near-circular and aligned orbits with orbital elements and angles randomized in the same manner as our two-planet simulations in \S \ref{sec:twobody}. The initial semimajor axes of the orbits were again informed by equating the initial orbital energy to that of the present-day system. Simulations were run for $10^9$ years. The planets were each separated by $4$ MHR, a somewhat arbitrary choice designed to induce rapid dynamical instability. Numerous studies \citep[e.g.][]{Chambers_1996, chatterjee_2008, anderson_2020} demonstrate that varying the orbital spacing between planets impacts only the time to instability in scattering simulations, not the final dynamical outcomes. To verify this, we have run identical ensembles with planets separated by $3$ MHR, with consistent results. We report here the results from the $4$ MHR separated cases.

\begin{table*}
    \begin{center}
    \begin{tabularx}{0.89\textwidth}{ll|cccc|r}
    \toprule
    Suite & Extra Planet Mass & $\geq 2$ Planets & Consistent Masses & $2\sigma$ Orbits & $1\sigma$ Orbits & Mean Energy Error\\
    \hline
    \texttt{suite0} & $0.05 \times \min(m_\mathrm{b}, m_\mathrm{c})$ & 4628 & 4253 & 0 & 0 & 0.08\\
    \texttt{suite1} & $0.25 \times \min(m_\mathrm{b}, m_\mathrm{c})$ & 5782 & 4855 & 139 & 0 & 0.15\\
    \texttt{suite2} & $0.5 \times \min(m_\mathrm{b}, m_\mathrm{c})$ & 5634 & 3497 & 39 & 0 & 0.25\\
    \texttt{suite3} & $\min(m_\mathrm{b}, m_\mathrm{c})$ & 5396 & 4942 & 19 & 2 & 0.41\\
    \texttt{suite4} & $\max(m_\mathrm{b}, m_\mathrm{c})$ & 5791 & 5183 & 22 & 4 & 0.37\\
    \texttt{suite5} & $1.5 \times \max(m_\mathrm{b}, m_\mathrm{c})$ & 6834 & 441 & 1 & 0 & 0.16\\
    \bottomrule
    \end{tabularx}
    \end{center}
    \caption{Compiled results from scattering simulations with three initial planets: 14 Her b and 14 Her c as well as an additional planet that may be ejected from the system. The first and second columns denote the name of the simulation suite and the mass of the additional plant added in that suite, respectively. The third column reports the number of simulations in each suite that are able to retain two bound planets. The fourth column reports the number of simulations in which the innermost two planets (often the only planets remaining) correctly correspond to 14 Her b and 14 Her c. The fifth and sixth columns report the number of simulations that match the observational constraints to $2\sigma$ and $1\sigma$ significance, respectively. These two columns only consider a match to the prograde peak of the posteriors, because no simulations across any suite matches the retrograde peak at $2\sigma$ confidence. The last column reports the mean relative energy error, $|\Delta E / E_0|$. Our results favor scenarios in which a planet either slightly less massive or around the mass of 14 Her b or c initially existed in the system and was ejected.}
    \label{tab:threebody}
\end{table*}
Table \ref{tab:threebody} compiles results from our simulations. We assessed each simulation's qualitative consistency with the observed system with increasing demands of rigor. First we checked if each simulation is able to retain two or more bound planets at all. The majority of simulations across all six suites satisfy this condition. Next, we checked if the two innermost bound planets correctly correspond to 14 Her b and 14 Her c, by discarding simulations in which the planet's masses fall outside of their $2\sigma$ confidence intervals. This disqualified the majority of simulations (around $95\%$) in \texttt{suite5}, in which the extra planet is more massive than both 14 Her b and 14 Her c. This is consistent with past results from the literature \citep[e.g.][]{rice2003substellar, raymond2008mmrs, marzari2025planet}, who showed that less massive planets tend to be preferentially ejected. 

Finally, we checked if the joint properties of the innermost two planets qualitatively satisfy the observational constraints with the same process as in \S \ref{sec:twobody}. No simulations in any suite reproduce the retrograde family of solutions at $2\sigma$ confidence. The values reported in Table \ref{tab:threebody} hence reflect matches to the prograde peak only. Our results favor the extra planet being less massive than either 14 Her b or c, with \texttt{suite1} providing the most matches. \texttt{suite0}, which features the smallest additional planet, does not generate sufficiently violent dynamical instabilities to reproduce the 14 Her system in any simulation, demonstrating that another very massive body is required in this scenario.

\begin{figure*}
    \centering
    \includegraphics{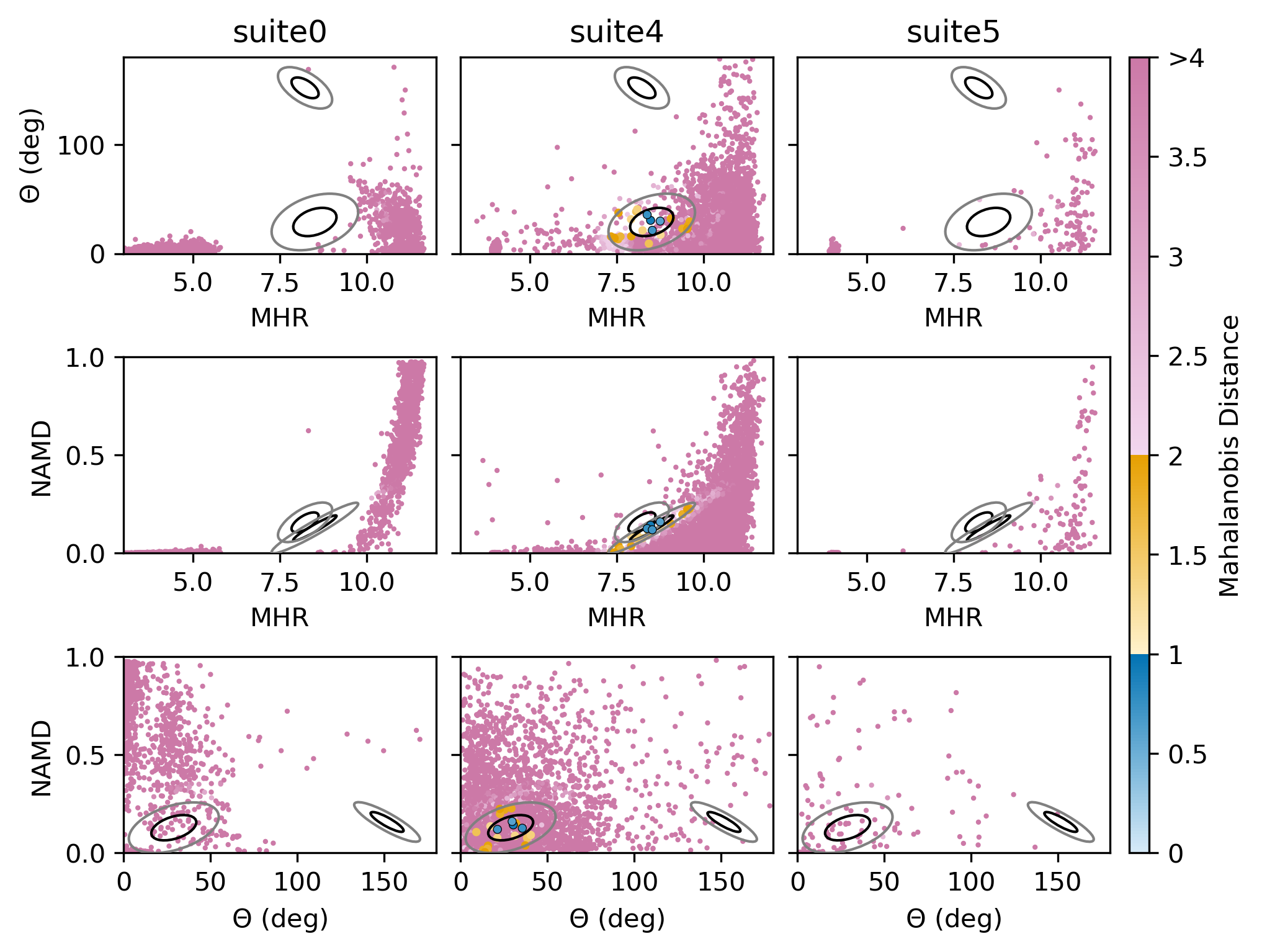}
    \caption{Simulations that retain a 14 Her b and 14 Her c analogue from \texttt{suite0}, \texttt{suite4} and \texttt{suite5}, plotted in mutual Hill radius (MHR) - mutual inclination ($\Theta$) - NAMD space of the two innermost planets. The $1\sigma$ and $2\sigma$ observational constraints are plotted as the black and gray ellipses, respectively. The colorbar corresponds to the minimum Mahalanobis distance $D_\mathrm{M}$ of the simulation to either the prograde or retrograde peak -- blue for systems that are a $1\sigma$ match to the data, orange for those that are $2\sigma$ matches, and pink otherwise. No simulations, across any suite, match the retrograde peak at $2\sigma$ or better significance. }
    \label{fig:3body_results}
\end{figure*}

Results from simulations in \texttt{suite0}, \texttt{suite4} and \texttt{suite5} that retain both a 14 Her b and 14 Her c analogue (corresponding to the fourth column in Table \ref{tab:threebody}) are plotted in Figure \ref{fig:3body_results}. The suites with an extra planet that was an appreciable fraction of the mass of the present-day planets, \texttt{suite1} - \texttt{suite4}, returned broadly consistent results. The suites with an equal-mass extra planet, \texttt{suite3} and \texttt{suite4}, tended to produce slightly more widely-spaced and more highly inclined systems than the suites with a less massive extra planet, \texttt{suite1} and \texttt{suite2}. Forming 14 Her is feasible in all four of these suites. The prospect of forming the system appears less feasible if the additional planet is more massive, as in \texttt{suite5}, or significantly less massive, as in \texttt{suite0}. In the rare cases in which the system does eject the more massive planet instead of 14 Her b or c, the resulting two-body system is invariably too widely separated. We note that in each simulation suite, there exists a small population of low eccentricity, low mutual inclination systems clustered around mutual separations near $4$ MHR. These are likely systems that did not undergo any significant instability over the $10^9$ year integration window, and are few enough in number that they do not impact our population-level results.

\begin{figure}
    \centering
    \includegraphics{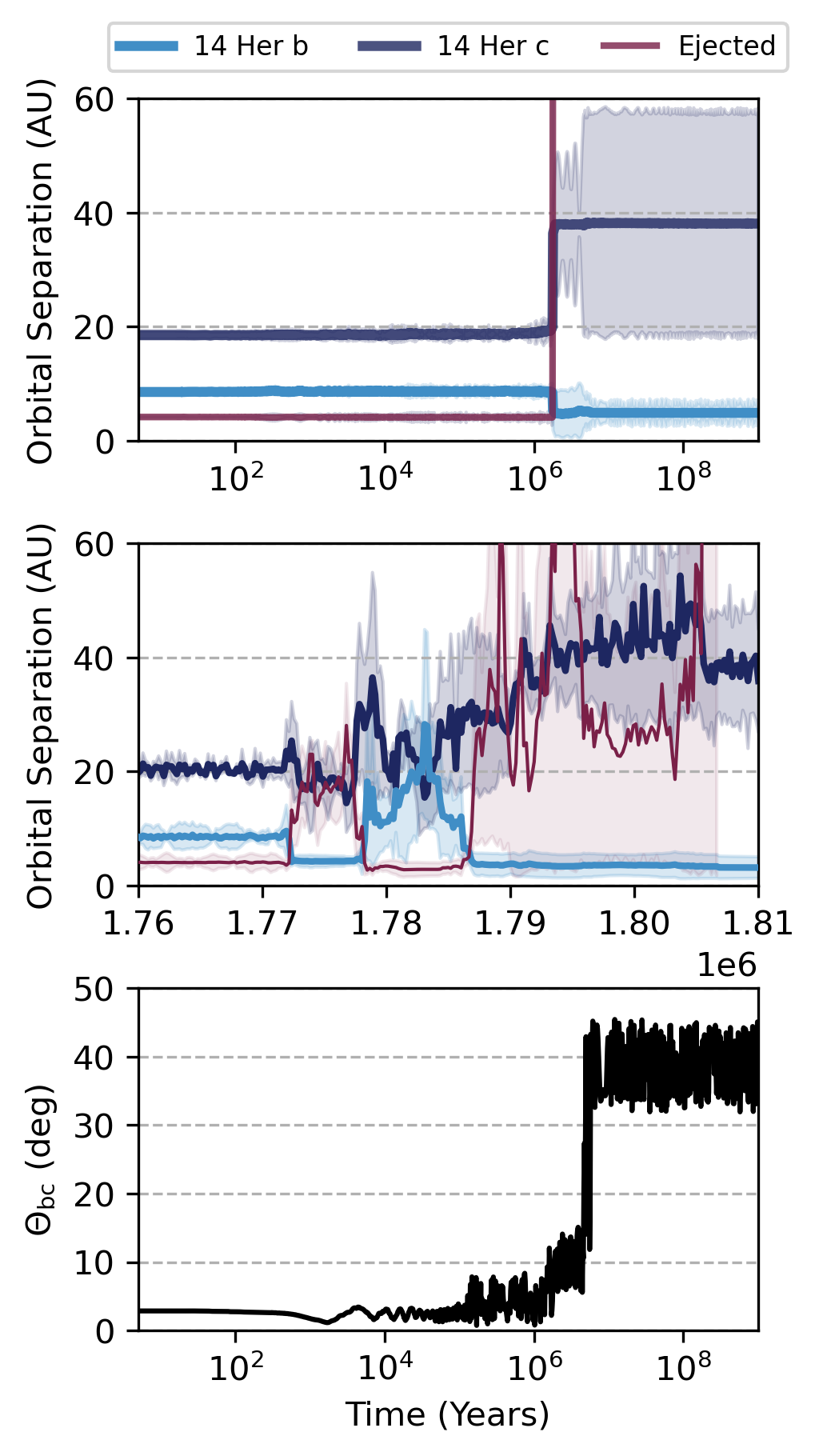}
    \caption{One simulation that successfully reproduces the present-day state of 14 Her. \textbf{Top}: Orbital separation of the three planets over the full integration window. Semimajor axes are plotted in solid lines, while the bounds of aphelion and perihelion are represented with the shaded contours. At this scale, the scattering event appear to happen nearly instantaneously. \textbf{Middle}: Zoom in between 1.76 and 1.81 Myr, where the scattering event takes place. The extra planet in the system is ejected after multiple close encounters between 14 Her b and 14 Her c cause an orbit crossing event between 14 Her b and the ejected planet. \textbf{Bottom}: Evolution of the mutual inclination between 14 Her b and 14 Her c. There is a good qualitative match to 14 Her's architecture.}
    \label{fig:example_evol}
\end{figure}

Figure \ref{fig:example_evol} shows the time-evolution of one system that successfully reproduces the present-day 14 Her system. This simulation is from \texttt{suite4}, and was initialized with $M_* = 1.004 M_\odot$, $m_\mathrm{extra} = 6.42 M_\mathrm{J}$, $m_\mathrm{b} = 9.89 M_\mathrm{J}$, and $m_\mathrm{c} = 9.89 M_\mathrm{J}$. The initial semimajor axes are $a_\mathrm{extra,i} = 4.15$ au, $a_\mathrm{b,i} = 8.54$ au, and $a_\mathrm{c,i} = 18.52$ au. All three planets remained on relatively circular, coplanar orbits for around 1.7 Myr. At this point, significant eccentricity oscillations were generated and 14 Her b and 14 Her c underwent multiple close encounters. The extra planet was ejected after a particularly violent close encounter that creates an orbit crossing event between 14 Her c and the ejected planet. This moment corresponded to a jump in the mutual inclination between 14 Her b and 14 Her c. The simulation is a qualitative match to the data to within $1\sigma$, with $D_\mathrm{M} = 0.739$.

We conclude that it is possible to generate the present-day mutually inclined and eccentric architecture of 14 Her via three-planet scattering in which one planet is ultimately ejected. The mass of the ejected planet must be a significant fraction of that of the planets currently residing in the system. The prograde family of solutions is reliably reproduced, while the retrograde solution is not. Our simulations demonstrate that a system with relatively low mutual inclination, which had ejected a less massive (of order a quarter to the mass of 14 Her b or 14 Her c) planet in its dynamical past, well reproduces the present-day 14 Her system architecture, though we note that other sets of initial conditions (such as starting out with more massive bodies) are also feasible and may lead to qualitatively different dynamical outcomes.

\section{Present-Day Secular Evolution}
\label{sec:present}

While the scattering simulations build our understanding of the formation history of 14 Her, we still have an incomplete picture of the system's current dynamic state. Exploring the present and future orbital evolution is also important to glean insight into the secular evolution of the eccentric and inclined orbits. Moreover, this investigation could also potentially rule out some unphysical parameter space of the posterior distribution. In particular, there may be a portion of posterior parameter space in which the inner planet undergoes eccentricity oscillations of a sufficient magnitude to trigger long-range high-eccentricity migration. Such system configurations could be ruled out, given that 14 Her b is a cold giant planet and has not become a hot Jupiter.  

Here we construct $N$-body simulations similar to those in \S \ref{sec:stab}. We consider the $8028$ stable system configurations from \S \ref{sec:stab} and again use the \texttt{WHFast} integrator with a timestep equal to 1/20th the effective period at pericenter of 14 Her b. In addition, we now incorporate tidal dissipation using an equilibrium tides model with the \texttt{tides\_spin} module \citep{Lu_2023} in \texttt{REBOUNDx} \citep{tamayo_reboundx}. 14 Her b is the only body whose tidal properties are relevant to the calculation. We estimate 14 Her b's radius to be $R_\mathrm{J}$ and its Love number to be $k_2 = 0.5$. We set its planetary tidal quality factor equal to $Q=10^4$, which we use to convert to a constant tidal time lag using $\tau=1/(2 n Q)$.  While this value of $Q$ is likely smaller than realistic values (which are probably closer to $Q\approx10^6$ for a gas giant planet), we use this smaller tidal quality factor to reduce our computation time. In effect, scaling $Q$ is equivalent to scaling time, such that our 10 Myr simulations are analogous to a 1 Gyr simulation if we had used $Q=10^6$. 

Analyzing the evolution in 14 Her b's semi-major axis, we do not find any simulations in which the inner planet undergoes tidal migration. This is because the planet's pericenter distance does not reach sufficiently small separations during eccentricity oscillations; the smallest it gets is ${\sim}0.3$ au, but most cases are much larger. These results imply that no stable system configurations are consistent with the inner planet undergoing high-eccentricity migration, so the posterior parameter space cannot be constrained further based on this argument.

\begin{figure}
    \centering
    \includegraphics[width=\columnwidth]{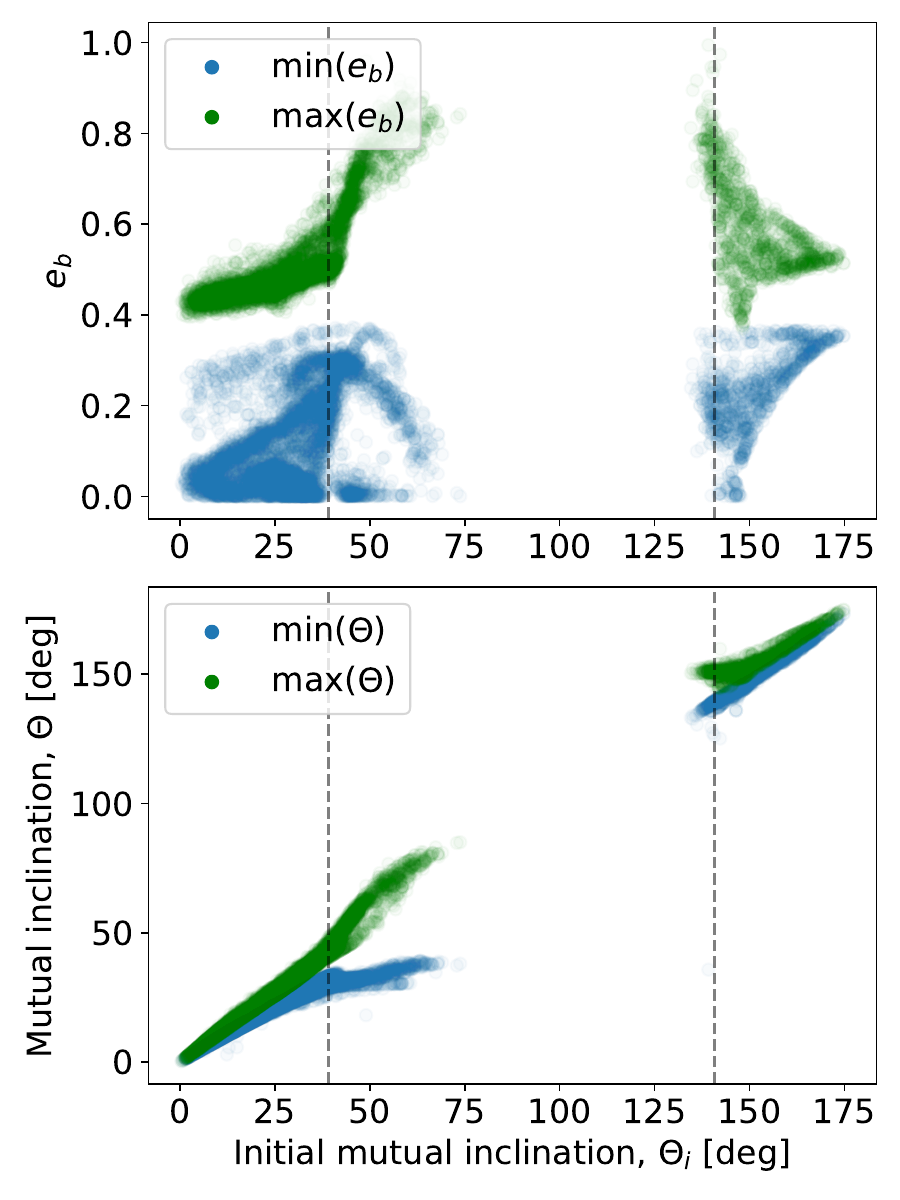}
    \caption{Summary of $8028$ $N$-body simulations representing the system's present-day evolution, corresponding to the subset of stable posteriors derived in \S \ref{sec:stab}. \textbf{Top}: 14 Her b's minimum and maximum eccentricity across each integration, plotted as a function of the initial mutual inclination. \textbf{Bottom}: Minimum and maximum mutual inclination across each simulation, plotted as a function of the initial mutual inclination. The dashed lines are at the nominal ``Kozai angles'', $39.2^{\circ}$ and $140.7^{\circ}$.  }
    \label{fig:present-day}
\end{figure}

While no configurations are consistent with tidal migration, many show extreme secular oscillations. Figure \ref{fig:present-day} shows the eccentricity and mutual inclination extremes. For each simulation, we compute the minimum and maximum eccentricity of planet b and the minimum and maximum mutual inclination between the two planets across the integration. We then plot these as a function of the initial mutual inclination, shown in Figure \ref{fig:present-day}. When the initial mutual inclination is between the nominal ``Kozai angles'', $39.2^\circ < \Theta < 140.7^\circ$, 14 Her b experiences large eccentricity and inclination oscillations. The eccentricity reaches $e_b\sim0.9$ at maximum, while the variations in the mutual inclination reach $\Delta\Theta\sim50^{\circ}$. Thus, depending on where the true system parameters lie, 14 Her may indeed be undergoing strong dynamical interactions while staying stable in the long term.

\section{Discussion}
\label{sec:disc}

\subsection{Stellar Flybys}
The assertion that an ejected planet is necessary to generate 14 Her's architecture rests upon the assumption that the architecture was generated via planet-planet scattering. This may not be the case. One alternative is a stellar flyby, which was briefly explored as the culprit for 14 Her's architecture by \cite{bgagliuffi2021her}, and could have induced dynamical instability in a system originally consisting of just two otherwise Hill stable planets. While possible, it is quite difficult for a flyby to excite significant eccentricities $(e \gtrsim 0.1)$. Authors such as \cite{zakamska2004flyby}, \cite{malmberg2011flyby} and \cite{li2020flyby} have demonstrated that violent dynamics of this magnitude are expected only for flybys with close approaches of order the semimajor axis of the outermost planet. There is no strong reason to expect the 14 Her system to have experienced such a close stellar encounter. Adopting representative local stellar densities ($n_* \sim0.1$ pc$^{-3}$, \citealt{bovy2017stellar}) and galactic field encounter speeds ($50$ km/s, \citealt{zakamska2004flyby}), the probability of a close ($\lesssim$ 50 AU) flyby over the few Gyr system lifetime is below 0.1\%. A more detailed kinematic analysis of likely flybys in the 14 Her system is warranted but beyond the scope of this work. We hence disfavor the prospect of a flyby that renders an ejection unnecessary.

\subsection{Formation Implications}
Our simulations imply a large mass budget for the planets initially in the 14 Her system. In addition to the two existing super-Jupiters, the demands of another ejected super-Jupiter imply a total initial planetary mass of $10-30M_J$.

This is not entirely unexpected, given 14 Her's very high metallicity of [Fe/H] = 0.43 \citep{soubiran2010pastel}. Well-known correlations exist between host star metallicity and giant planet occurrence rate \citep{fischer2005planet}, as well as between giant planet metallicity and mass \citep{thorngren2016mass, chachan2025mass}. In addition, tentative evidence has emerged supporting a correlation between host star metallicity and total planetary mass budget in recent years: \cite{nguyen2025metallicity} and \cite{howe2026host} both showed that systems with multiple giant planets are more common around metal-rich stars. Furthermore, \cite{nguyen2025metallicity} draw a direct link between total planetary mass and stellar metallicity and assert that a system with 14 Her's metallicity may be expected to host upwards of 50 $M_\mathrm{J}$ worth of planetary mass at once.

An implication of the correlation between total planetary mass and metallicity is that planetary systems around metal-rich stars may be more dynamically active, with more planetary mass budget to gravitationally interact with. This was first posited by \cite{dawson2013giant}, who noted that the most eccentric giant planets orbit metal-rich stars. In this framework, it is not unreasonable to assume a metal-rich star such as 14 Her may have been able to host three or more super-Jupiters.

\subsection{Insights From Ejection Statistics}
\begin{figure}
    \centering
    \includegraphics[width=0.48\textwidth]{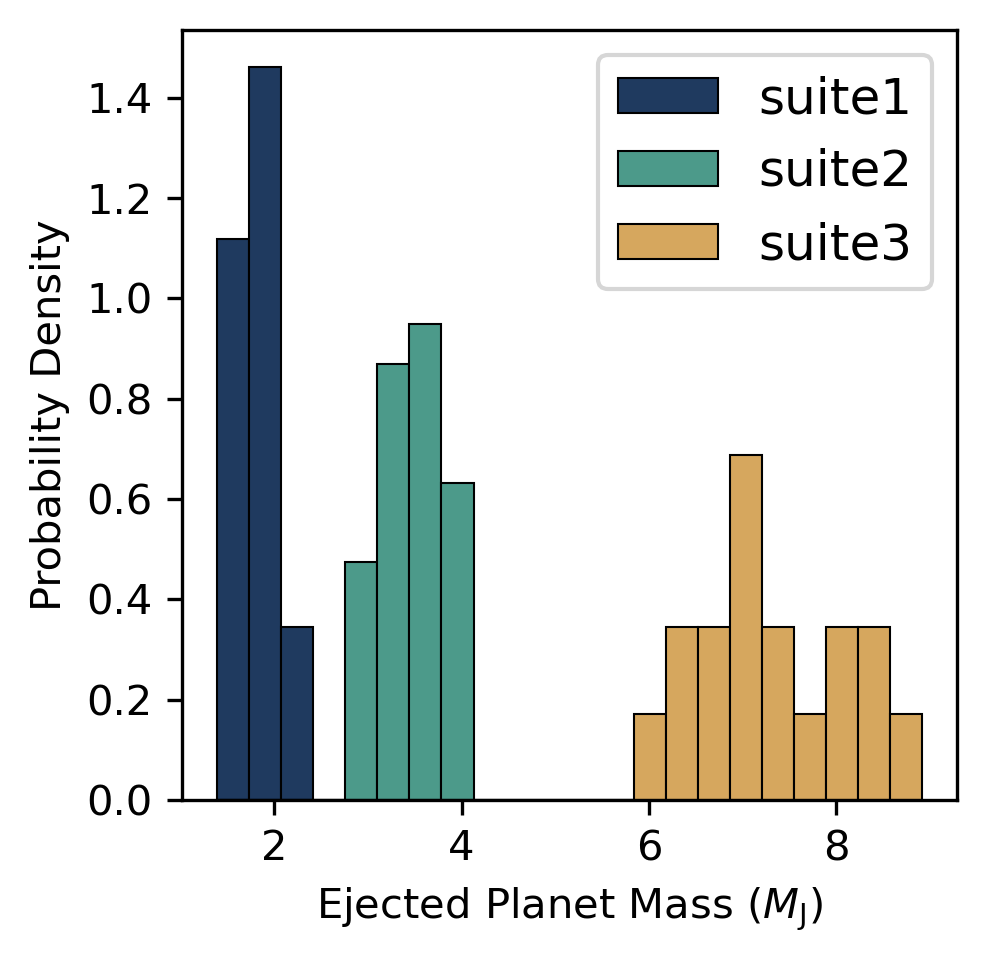}
    \caption{Histograms of the mass of the ejected planet for simulations which reproduce 14 Her's architecture at the $2\sigma$ level. We have plotted results from \texttt{suite1} (blue), \texttt{suite2} (teal), and \texttt{suite3} (yellow). \texttt{suite0} and \texttt{suite5} both fail to record enough 14 Her analogues, and \texttt{suite4} is qualitatively similar to \texttt{suite3} and thus omitted for clarity. Varying initial system architectures result in clearly distinct ejection statistics.}
    \label{fig:ejection_stats}
\end{figure}

The vast majority of our scattering simulations require a planet of comparable mass to 14 Her b or c to be ejected from the system. Figure \ref{fig:ejection_stats} reports statistics for the masses of the ejected planet necessary to reproduce the 14 Her system at the $2\sigma$ level. Each suite predicts a distinct distribution of ejected planet masses, as the most massive planets in each system are expected to be preferentially retained.

It is worth emphasizing that the unusual nature of the 14 Her system is only apparent once the outer companion is taken into account. Had 14 Her c remained undetected, 14 Her b would appear to be a fairly typical eccentric super-Jupiter, residing near the peak of the eccentricity distribution of cold giant planets at $(e \sim 0.3)$ \citep{blunt2026eccentricity}. A variety of mechanisms have been proposed to explain the eccentricities of single giant planets, including planet-disk interactions \citep[e.g.][]{lega2021eccentricity} and secular perturbations from unseen stellar companions \citep{weldon2025cold}. However, the discovery of 14 Her c reveals that the system is not simply an eccentric single-planet system, but rather a pair of massive planets on eccentric and significantly misaligned orbits. This architecture is difficult to reconcile with mechanisms that act primarily on a single planet and instead points more naturally towards the scattering scenario explored in this work. More planets discovered in the parameter space of 14 Her c will illuminate the relative importance of these mechanisms in sculpting the architectures of cold giant systems.

The architectures of the outer reaches of planetary systems are largely shrouded in mystery. For instance, do outer gas giants exhibit the same intriguing intra-system mass uniformity observed in the Kepler multis \citep{millholland2017peas, weiss2018peas}, or will they exhibit a wider range of masses as seen in our own outer Solar System? Will they be in dynamically hot configurations as 14 Her is, or rather in pristine resonant chains such as HR 8799 \citep{wang2018hd}? Our study highlights that valuable information regarding initial conditions of outer giants may be gleaned from both the current system state and ultimately ejected planets. 

Our simulations only considered the case of a single massive ejected planet. In principle, numerous scattering events with lower-mass companions could have also imparted cumulative recoil kicks to a giant planet to generate significant eccentricities and subsequent dynamical instabilities, as seen in planetesimal-driven orbital evolution \citep[e.g.][]{hahn1999planetesimal}. This scenario would naturally change the demographics of predicted ejected planets. While further investigation into this specific scenario is warranted, we judge scattering of multiple smaller planets as a less likely progenitor as eccentricity evolution is often diffusive in the limit of much smaller perturbers \citep[e.g.][]{kirsh2009planetesimal}.

The possibility that the present-day architecture of 14 Her encodes evidence for a previously ejected giant planet is not unique. Similar conclusions have been drawn for several other systems, including $\upsilon$ Andromedae \citep{ford2005ups} and GJ 1148 \citep{yuan2024gj1148}. In this sense, 14 Her joins a small but growing class of systems whose present-day orbital architectures may preserve evidence for primordial giant planet ejections. An intriguing consequence of these interpretations is that the ejected perturber would fall well within the sensitivity range of Roman's microlensing survey \citep{johnson2020ffp}.

In the near future, the fourth and fifth data releases of the \textit{Gaia} mission are projected to discover ${\sim}38,000$ giant planets on orbital separations from $2-5$ au \citep{lammers2025exoplanet}, while the Nancy Grace Roman Telescope is expected to discover ${\sim}1000$ free floating planets. Taken together these will offer large-scale population-level insights into the outer reaches of planetary systems, and we speculate that powerful constraints will be possible for initial system architectures on a population level.

\section{Conclusions}
\label{sec:conclusions}
The 14 Her system, consisting of two eccentric misaligned super-Jupiters, is a pristine relic of the dynamical mechanisms that are believed to shape many exoplanetary systems. We have extensively analyzed the system's dynamical history. Our main conclusions are summarized below.

\begin{itemize}
    \item Planet-planet scattering is able to satisfactorily qualitatively reproduce all key aspects of 14 Her's architecture, \textbf{only} if at least one other massive body originally existed in the system and was since ejected. Variations in the mass of this ejected planet can be distinctly attributed to different initial system configurations. We hence speculate that upcoming results from Roman's microlensing survey may shed light on the initial architectures of outer giant planet systems.
    \item None of our simulations are consistent with the retrograde family of solutions identified by \cta{bgagliuffi2025jwst}. The observational degeneracy responsible for the bimodal distributions of mutual inclination will be broken with the release of time-series astrometric data from \textit{Gaia} DR4. If it is indeed revealed that 14 Her occupies this retrograde region of parameter space, an alternative story will be necessary to explain its architecture.
    \item From an analysis of the system's long-term stability, we strongly disfavor mutual inclinations between $75-105^\circ$ and slightly disfavor mutual inclinations between $39.2-140^\circ$. Our results suggest that the 14 Her system is, at present-day, not experiencing traditional ZLK oscillations. We note, however, that ZLK oscillations are not strictly defined in the regime of two roughly equal-mass particles \citep{naoz2016eccentric}, and that the 14 Her system is nonetheless likely still experiencing significant eccentricity and inclination oscillations. However, none of the stable posteriors are consistent with high-eccentricity tidal migration as the pericenter distance of 14 Her b does not reach sufficiently small values.
\end{itemize}
Considerable observing time and targeted effort was required to characterize the orbital architecture of 14 Her. In the era of \textit{Gaia} DR4/DR5 and \textit{Roman}, this may not be the case for similar systems in the near future. A single case study, while illuminating, cannot rule out entire formation pathways. While our simulations strongly suggest that $N=2$ planet scattering cannot be the progenitor of 14 Her, it is possible that this scenario is feasible but simply rare. If many more such systems are discovered, such formation pathways will be possible to rule out from statistical and demographic arguments. 

Early intriguing hints have already emerged: RV follow-up of the Gaia-4b and Gaia-5b planet candidates revealed super-Jupiters/brown dwarfs on eccentric orbits\footnote{Though \cite{yahalomi2026astrometric} cautions that circular planet pairs in 2:1 mean motion resonance may masquerade as single eccentric planets in \textit{Gaia} data, which would represent a very different dynamical interpretation.} \citep{stefansson2025gaia} reminiscent of a scattering population which would have similarly resulted in massive primordial ejections. If the processes responsible for shaping 14 Her's architecture are indeed representative of population-level trends, we may expect the upcoming Data Releases of the \textit{Gaia} mission to uncover many more eccentric, misaligned planets.

\begin{acknowledgments}
We thank the reviewer Max Goldberg for a constructive and helpful report. We also thank Jiayin Dong, Sam Hadden, Yubo Su, Dan Tamayo, Yixian Chen, Fritz Benedict, the members of the CCA Astronomical Data Group and the Rice Research Group for insightful conversations. This work has benefited from use of the \texttt{GRACE} computing cluster at the Yale Center for Research Computing and the \texttt{rusty} cluster at the Flatiron Institute. T.L.\ is supported by a Flatiron Research Fellowship at the Flatiron Institute, a division of the Simons Foundation. M.R. acknowledges support from Heising-Simons Foundation Grants \#2021-2802 and \#2023-4478, as well as NASA Exoplanets Research Program NNH23ZDA001N-XRP (grant No. 80NSSC24K0153). The analysis performed in this work greatly benefited from the \texttt{SimulationArchive} feature in \texttt{REBOUND} \citep{rein2017simarchive}.
\end{acknowledgments}

\software{\texttt{corner} \citep{dfm2016corner}, \texttt{matplotlib} \citep{hunter_2007}, \texttt{numpy} \citep{oliphantnumpy, vanderwalt2011numpy} \texttt{REBOUND} \citep{Rein_2012}, \texttt{REBOUNDx} \citep{tamayo_reboundx}, \texttt{scipy} \citep{virtanen2020scipy}, \texttt{slurm} \citep{jette2002slurm}}

\bibliography{sample7}{}

\begin{thebibliography}{}
\expandafter\ifx\csname natexlab\endcsname\relax\def\natexlab#1{#1}\fi
\providecommand{\url}[1]{\href{#1}{#1}}
\providecommand{\dodoi}[1]{doi:~\href{http://doi.org/#1}{\nolinkurl{#1}}}
\providecommand{\doeprint}[1]{\href{http://ascl.net/#1}{\nolinkurl{http://ascl.net/#1}}}
\providecommand{\doarXiv}[1]{\href{https://arxiv.org/abs/#1}{\nolinkurl{https://arxiv.org/abs/#1}}}

\bibitem[{Q. {An} {et~al.}(2025){An}, {Lu}, {Brandt}, {Brandt}, \& {Li}}]{An2025significant}
{An}, Q., {Lu}, T., {Brandt}, G.~M., {Brandt}, T.~D., \& {Li}, G. 2025, \bibinfo{title}{{Significant Mutual Inclinations Between the Stellar Spin and the Orbits of Both Planets in the HAT-P-11 System},} \aj, 169, 22, \dodoi{10.3847/1538-3881/ad90b4}

\bibitem[{K.~R. {Anderson} {et~al.}(2020){Anderson}, {Lai}, \& {Pu}}]{anderson_2020}
{Anderson}, K.~R., {Lai}, D., \& {Pu}, B. 2020, \bibinfo{title}{{In situ scattering of warm Jupiters and implications for dynamical histories},} \mnras, 491, 1369, \dodoi{10.1093/mnras/stz3119}

\bibitem[{T.~W. Anderson {et~al.}(1958)Anderson, Anderson, Anderson, Anderson, \& Math{\'e}maticien}]{anderson1958introduction}
Anderson, T.~W., Anderson, T.~W., Anderson, T.~W., Anderson, T.~W., \& Math{\'e}maticien, E.-U. 1958, An introduction to multivariate statistical analysis, Vol.~2 (Wiley New York)

\bibitem[{P.~J. {Armitage}(2010){Armitage}}]{armitage2010pf}
{Armitage}, P.~J. 2010, {Astrophysics of Planet Formation}

\bibitem[{D.~C. {Bardalez Gagliuffi} {et~al.}(2021){Bardalez Gagliuffi}, {Faherty}, {Li}, {Brandt}, {Williams}, {Brandt}, \& {Gelino}}]{bgagliuffi2021her}
{Bardalez Gagliuffi}, D.~C., {Faherty}, J.~K., {Li}, Y., {et~al.} 2021, \bibinfo{title}{{14 Her: A Likely Case of Planet-Planet Scattering},} \apjl, 922, L43, \dodoi{10.3847/2041-8213/ac382c}

\bibitem[{D.~C. {Bardalez Gagliuffi} {et~al.}(2025){Bardalez Gagliuffi}, {Balmer}, {Pueyo}, {Brandt}, {Giovinazzi}, {Millholland}, {Black}, {Lu}, {Rice}, {Mang}, {Morley}, {Lacy}, {Girard}, {Matthews}, {Carter}, {Bowler}, {Faherty}, {Fontanive}, \& {Rickman}}]{bgagliuffi2025jwst}
{Bardalez Gagliuffi}, D.~C., {Balmer}, W.~O., {Pueyo}, L., {et~al.} 2025, \bibinfo{title}{{JWST Coronagraphic Images of 14 Her c: A Cold Giant Planet in a Dynamically Hot Multiplanet System},} \apjl, 988, L18, \dodoi{10.3847/2041-8213/ade30f}

\bibitem[{C. {Beaug{\'e}} \& D. {Nesvorn{\'y}}(2012){Beaug{\'e}} \& {Nesvorn{\'y}}}]{beauge2012multiple}
{Beaug{\'e}}, C., \& {Nesvorn{\'y}}, D. 2012, \bibinfo{title}{{Multiple-planet Scattering and the Origin of Hot Jupiters},} \apj, 751, 119, \dodoi{10.1088/0004-637X/751/2/119}

\bibitem[{G.~F. {Benedict} {et~al.}(2023){Benedict}, {McArthur}, {Nelan}, \& {Bean}}]{benedict2023her}
{Benedict}, G.~F., {McArthur}, B.~E., {Nelan}, E.~P., \& {Bean}, J.~L. 2023, \bibinfo{title}{{The 14 Her Planetary System: Companion Masses and Architecture from Radial Velocities and Astrometry},} \aj, 166, 27, \dodoi{10.3847/1538-3881/acd93a}

\bibitem[{H.~G. {Bhaskar} \& H. {Perets}(2024){Bhaskar} \& {Perets}}]{bhaskar2024dynamical}
{Bhaskar}, H.~G., \& {Perets}, H. 2024, \bibinfo{title}{{Dynamical and Secular Stability of Mutually Inclined Planetary Systems},} \apj, 973, 108, \dodoi{10.3847/1538-4357/ad62f9}

\bibitem[{S. {Blunt} {et~al.}(2026){Blunt}, {Wang}, {Murray-Clay}, {Macintosh}, {Rubenzahl}, \& {Fulton}}]{blunt2026eccentricity}
{Blunt}, S., {Wang}, J., {Murray-Clay}, R., {et~al.} 2026, \bibinfo{title}{{Evidence for a Peak at {\ensuremath{\sim}}0.3 in the Eccentricity Distribution of Typical Super-Jovian Exoplanets},} \apjl, 999, L26, \dodoi{10.3847/2041-8213/ae41c3}

\bibitem[{W.~J. {Borucki} {et~al.}(2010){Borucki}, {Koch}, {Basri}, {Batalha}, {Brown}, {Caldwell}, {Caldwell}, {Christensen-Dalsgaard}, {Cochran}, {DeVore}, {Dunham}, {Dupree}, {Gautier}, {Geary}, {Gilliland}, {Gould}, {Howell}, {Jenkins}, {Kondo}, {Latham}, {Marcy}, {Meibom}, {Kjeldsen}, {Lissauer}, {Monet}, {Morrison}, {Sasselov}, {Tarter}, {Boss}, {Brownlee}, {Owen}, {Buzasi}, {Charbonneau}, {Doyle}, {Fortney}, {Ford}, {Holman}, {Seager}, {Steffen}, {Welsh}, {Rowe}, {Anderson}, {Buchhave}, {Ciardi}, {Walkowicz}, {Sherry}, {Horch}, {Isaacson}, {Everett}, {Fischer}, {Torres}, {Johnson}, {Endl}, {MacQueen}, {Bryson}, {Dotson}, {Haas}, {Kolodziejczak}, {Van Cleve}, {Chandrasekaran}, {Twicken}, {Quintana}, {Clarke}, {Allen}, {Li}, {Wu}, {Tenenbaum}, {Verner}, {Bruhweiler}, {Barnes}, \& {Prsa}}]{borucki_2010}
{Borucki}, W.~J., {Koch}, D., {Basri}, G., {et~al.} 2010, \bibinfo{title}{{Kepler Planet-Detection Mission: Introduction and First Results},} Science, 327, 977, \dodoi{10.1126/science.1185402}

\bibitem[{J. {Bovy}(2017){Bovy}}]{bovy2017stellar}
{Bovy}, J. 2017, \bibinfo{title}{{Stellar inventory of the solar neighbourhood using Gaia DR1},} \mnras, 470, 1360, \dodoi{10.1093/mnras/stx1277}

\bibitem[{T.~D. {Brandt}(2018){Brandt}}]{brandt2018hgca}
{Brandt}, T.~D. 2018, \bibinfo{title}{{The Hipparcos-Gaia Catalog of Accelerations},} \apjs, 239, 31, \dodoi{10.3847/1538-4365/aaec06}

\bibitem[{R.~P. {Butler} {et~al.}(2003){Butler}, {Marcy}, {Vogt}, {Fischer}, {Henry}, {Laughlin}, \& {Wright}}]{butler2003keck}
{Butler}, R.~P., {Marcy}, G.~W., {Vogt}, S.~S., {et~al.} 2003, \bibinfo{title}{{Seven New Keck Planets Orbiting G and K Dwarfs},} \apj, 582, 455, \dodoi{10.1086/344570}

\bibitem[{Y. {Chachan} {et~al.}(2025){Chachan}, {Fortney}, {Ohno}, {Thorngren}, \& {Murray-Clay}}]{chachan2025mass}
{Chachan}, Y., {Fortney}, J.~J., {Ohno}, K., {Thorngren}, D., \& {Murray-Clay}, R. 2025, \bibinfo{title}{{Revising the Giant Planet Mass─Metallicity Relation: Deciphering the Formation Sequence of Giant Planets},} \apj, 994, 43, \dodoi{10.3847/1538-4357/ae0cbf}

\bibitem[{J.~E. {Chambers} \& G.~W. {Wetherill}(1998){Chambers} \& {Wetherill}}]{chambers1998making}
{Chambers}, J.~E., \& {Wetherill}, G.~W. 1998, \bibinfo{title}{{Making the Terrestrial Planets: N-Body Integrations of Planetary Embryos in Three Dimensions},} \icarus, 136, 304, \dodoi{10.1006/icar.1998.6007}

\bibitem[{J.~E. {Chambers} {et~al.}(1996){Chambers}, {Wetherill}, \& {Boss}}]{Chambers_1996}
{Chambers}, J.~E., {Wetherill}, G.~W., \& {Boss}, A.~P. 1996, \bibinfo{title}{{The Stability of Multi-Planet Systems},} \icarus, 119, 261, \dodoi{10.1006/icar.1996.0019}

\bibitem[{S. {Chatterjee} {et~al.}(2008){Chatterjee}, {Ford}, {Matsumura}, \& {Rasio}}]{chatterjee_2008}
{Chatterjee}, S., {Ford}, E.~B., {Matsumura}, S., \& {Rasio}, F.~A. 2008, \bibinfo{title}{{Dynamical Outcomes of Planet-Planet Scattering},} \apj, 686, 580, \dodoi{10.1086/590227}

\bibitem[{J. {Chen} \& D. {Kipping}(2017){Chen} \& {Kipping}}]{chen2017radius}
{Chen}, J., \& {Kipping}, D. 2017, \bibinfo{title}{{Probabilistic Forecasting of the Masses and Radii of Other Worlds},} \apj, 834, 17, \dodoi{10.3847/1538-4357/834/1/17}

\bibitem[{P. {Cresswell} {et~al.}(2007){Cresswell}, {Dirksen}, {Kley}, \& {Nelson}}]{cresswell2007evolution}
{Cresswell}, P., {Dirksen}, G., {Kley}, W., \& {Nelson}, R.~P. 2007, \bibinfo{title}{{On the evolution of eccentric and inclined protoplanets embedded in protoplanetary disks},} \aap, 473, 329, \dodoi{10.1051/0004-6361:20077666}

\bibitem[{R.~I. {Dawson} \& R.~A. {Murray-Clay}(2013){Dawson} \& {Murray-Clay}}]{dawson2013giant}
{Dawson}, R.~I., \& {Murray-Clay}, R.~A. 2013, \bibinfo{title}{{Giant Planets Orbiting Metal-rich Stars Show Signatures of Planet-Planet Interactions},} \apjl, 767, L24, \dodoi{10.1088/2041-8205/767/2/L24}

\bibitem[{K.~M. {Deck} {et~al.}(2012){Deck}, {Holman}, {Agol}, {Carter}, {Lissauer}, {Ragozzine}, \& {Winn}}]{deck2012rapid}
{Deck}, K.~M., {Holman}, M.~J., {Agol}, E., {et~al.} 2012, \bibinfo{title}{{Rapid Dynamical Chaos in an Exoplanetary System},} \apjl, 755, L21, \dodoi{10.1088/2041-8205/755/1/L21}

\bibitem[{J. {Dong} {et~al.}(2026){Dong}, {Lee}, {Kokubo}, {Murray-Clay}, \& {Gupta}}]{dong2026scattering}
{Dong}, J., {Lee}, E.~J., {Kokubo}, E., {Murray-Clay}, R., \& {Gupta}, A. 2026, \bibinfo{title}{{Planet-Planet Scattering Explains the Mass-Eccentricity Relation of Warm Jupiters},} arXiv e-prints, arXiv:2603.22426, \dodoi{10.48550/arXiv.2603.22426}

\bibitem[{J. {Esposito} {et~al.}(2026){Esposito}, {Li}, \& {Wang}}]{esposito2026scattering}
{Esposito}, J., {Li}, G., \& {Wang}, S. 2026, \bibinfo{title}{{Unified Formation Channel of Hot and Warm Jupiters via Planet-Planet Scattering},} arXiv e-prints, arXiv:2603.22409, \dodoi{10.48550/arXiv.2603.22409}

\bibitem[{R.~B. {Fernandes} {et~al.}(2019){Fernandes}, {Mulders}, {Pascucci}, {Mordasini}, \& {Emsenhuber}}]{fernandes2019turnover}
{Fernandes}, R.~B., {Mulders}, G.~D., {Pascucci}, I., {Mordasini}, C., \& {Emsenhuber}, A. 2019, \bibinfo{title}{{Hints for a Turnover at the Snow Line in the Giant Planet Occurrence Rate},} \apj, 874, 81, \dodoi{10.3847/1538-4357/ab0300}

\bibitem[{D.~A. {Fischer} \& J. {Valenti}(2005){Fischer} \& {Valenti}}]{fischer2005planet}
{Fischer}, D.~A., \& {Valenti}, J. 2005, \bibinfo{title}{{The Planet-Metallicity Correlation},} \apj, 622, 1102, \dodoi{10.1086/428383}

\bibitem[{E.~B. {Ford} {et~al.}(2001){Ford}, {Havlickova}, \& {Rasio}}]{Ford_2001}
{Ford}, E.~B., {Havlickova}, M., \& {Rasio}, F.~A. 2001, \bibinfo{title}{{Dynamical Instabilities in Extrasolar Planetary Systems Containing Two Giant Planets},} \icarus, 150, 303, \dodoi{10.1006/icar.2001.6588}

\bibitem[{E.~B. {Ford} {et~al.}(2005){Ford}, {Lystad}, \& {Rasio}}]{ford2005ups}
{Ford}, E.~B., {Lystad}, V., \& {Rasio}, F.~A. 2005, \bibinfo{title}{{Planet-planet scattering in the upsilon Andromedae system},} \nat, 434, 873, \dodoi{10.1038/nature03427}

\bibitem[{E.~B. {Ford} \& F.~A. {Rasio}(2008){Ford} \& {Rasio}}]{ford2008}
{Ford}, E.~B., \& {Rasio}, F.~A. 2008, \bibinfo{title}{{Origins of Eccentric Extrasolar Planets: Testing the Planet-Planet Scattering Model},} \apj, 686, 621, \dodoi{10.1086/590926}

\bibitem[{D. Foreman-Mackey(2016)Foreman-Mackey}]{dfm2016corner}
Foreman-Mackey, D. 2016, \bibinfo{title}{corner.py: Scatterplot matrices in Python,} The Journal of Open Source Software, 1, 24, \dodoi{10.21105/joss.00024}

\bibitem[{B.~J. {Fulton} {et~al.}(2021){Fulton}, {Rosenthal}, {Hirsch}, {Isaacson}, {Howard}, {Dedrick}, {Sherstyuk}, {Blunt}, {Petigura}, {Knutson}, {Behmard}, {Chontos}, {Crepp}, {Crossfield}, {Dalba}, {Fischer}, {Henry}, {Kane}, {Kosiarek}, {Marcy}, {Rubenzahl}, {Weiss}, \& {Wright}}]{fulton2021cls}
{Fulton}, B.~J., {Rosenthal}, L.~J., {Hirsch}, L.~A., {et~al.} 2021, \bibinfo{title}{{California Legacy Survey. II. Occurrence of Giant Planets beyond the Ice Line},} \apjs, 255, 14, \dodoi{10.3847/1538-4365/abfcc1}

\bibitem[{ {\unskip Gaia Collaboration} {et~al.}(2023){\unskip Gaia Collaboration}, {Vallenari}, {Brown}, {Prusti}, {de Bruijne}, {Arenou}, {Babusiaux}, {Biermann}, {Creevey}, {Ducourant}, {Evans}, {Eyer}, {Guerra}, {Hutton}, {Jordi}, {Klioner}, {Lammers}, {Lindegren}, {Luri}, {Mignard}, {Panem}, {Pourbaix}, {Randich}, {Sartoretti}, {Soubiran}, {Tanga}, {Walton}, {Bailer-Jones}, {Bastian}, {Drimmel}, {Jansen}, {Katz}, {Lattanzi}, {van Leeuwen}, {Bakker}, {Cacciari}, {Casta{\~n}eda}, {De Angeli}, {Fabricius}, {Fouesneau}, {Fr{\'e}mat}, {Galluccio}, {Guerrier}, {Heiter}, {Masana}, {Messineo}, {Mowlavi}, {Nicolas}, {Nienartowicz}, {Pailler}, {Panuzzo}, {Riclet}, {Roux}, {Seabroke}, {Sordo}, {Th{\'e}venin}, {Gracia-Abril}, {Portell}, {Teyssier}, {Altmann}, {Andrae}, {Audard}, {Bellas-Velidis}, {Benson}, {Berthier}, {Blomme}, {Burgess}, {Busonero}, {Busso}, {C{\'a}novas}, {Carry}, {Cellino}, {Cheek}, {Clementini}, {Damerdji}, {Davidson}, {de Teodoro}, {Nu{\~n}ez Campos}, {Delchambre}, {Dell'Oro}, {Esquej},
  {Fern{\'a}ndez-Hern{\'a}ndez}, {Fraile}, {Garabato}, {Garc{\'\i}a-Lario}, {Gosset}, {Haigron}, {Halbwachs}, {Hambly}, {Harrison}, {Hern{\'a}ndez}, {Hestroffer}, {Hodgkin}, {Holl}, {Jan{\ss}en}, {Jevardat de Fombelle}, {Jordan}, {Krone-Martins}, {Lanzafame}, {L{\"o}ffler}, {Marchal}, {Marrese}, {Moitinho}, {Muinonen}, {Osborne}, {Pancino}, {Pauwels}, {Recio-Blanco}, {Reyl{\'e}}, {Riello}, {Rimoldini}, {Roegiers}, {Rybizki}, {Sarro}, {Siopis}, {Smith}, {Sozzetti}, {Utrilla}, {van Leeuwen}, {Abbas}, {{\'A}brah{\'a}m}, {Abreu Aramburu}, {Aerts}, {Aguado}, {Ajaj}, {Aldea-Montero}, {Altavilla}, {{\'A}lvarez}, {Alves}, {Anders}, {Anderson}, {Anglada Varela}, {Antoja}, {Baines}, {Baker}, {Balaguer-N{\'u}{\~n}ez}, {Balbinot}, {Balog}, {Barache}, {Barbato}, {Barros}, {Barstow}, {Bartolom{\'e}}, {Bassilana}, {Bauchet}, {Becciani}, {Bellazzini}, {Berihuete}, {Bernet}, {Bertone}, {Bianchi}, {Binnenfeld}, {Blanco-Cuaresma}, {Blazere}, {Boch}, {Bombrun}, {Bossini}, {Bouquillon}, {Bragaglia}, {Bramante}, {Breedt},
  {Bressan}, {Brouillet}, {Brugaletta}, {Bucciarelli}, {Burlacu}, {Butkevich}, {Buzzi}, {Caffau}, {Cancelliere}, {Cantat-Gaudin}, {Carballo}, {Carlucci}, {Carnerero}, {Carrasco}, {Casamiquela}, {Castellani}, {Castro-Ginard}, {Chaoul}, {Charlot}, {Chemin}, {Chiaramida}, {Chiavassa}, {Chornay}, {Comoretto}, {Contursi}, {Cooper}, {Cornez}, {Cowell}, {Crifo}, {Cropper}, {Crosta}, {Crowley}, {Dafonte}, {Dapergolas}, {David}, {David}, {de Laverny}, {De Luise}, \& {De March}}]{gaiadr3}
{\unskip Gaia Collaboration}, {Vallenari}, A., {Brown}, A.~G.~A., {et~al.} 2023, \bibinfo{title}{{Gaia Data Release 3. Summary of the content and survey properties},} \aap, 674, A1, \dodoi{10.1051/0004-6361/202243940}

\bibitem[{B. {Gladman}(1993){Gladman}}]{Gladman93}
{Gladman}, B. 1993, \bibinfo{title}{{Dynamics of Systems of Two Close Planets},} \icarus, 106, 247, \dodoi{10.1006/icar.1993.1169}

\bibitem[{P. {Goldreich} \& S. {Tremaine}(1980){Goldreich} \& {Tremaine}}]{goldreich1980disk}
{Goldreich}, P., \& {Tremaine}, S. 1980, \bibinfo{title}{{Disk-satellite interactions.},} \apj, 241, 425, \dodoi{10.1086/158356}

\bibitem[{S. {Hadden} \& Y. {Lithwick}(2018){Hadden} \& {Lithwick}}]{hadden_lithwick_2018}
{Hadden}, S., \& {Lithwick}, Y. 2018, \bibinfo{title}{{A Criterion for the Onset of Chaos in Systems of Two Eccentric Planets},} \aj, 156, 95, \dodoi{10.3847/1538-3881/aad32c}

\bibitem[{J.~M. {Hahn} \& R. {Malhotra}(1999){Hahn} \& {Malhotra}}]{hahn1999planetesimal}
{Hahn}, J.~M., \& {Malhotra}, R. 1999, \bibinfo{title}{{Orbital Evolution of Planets Embedded in a Planetesimal Disk},} \aj, 117, 3041, \dodoi{10.1086/300891}

\bibitem[{D.~M. {Hernandez} {et~al.}(2022){Hernandez}, {Zeebe}, \& {Hadden}}]{hernandez_2022}
{Hernandez}, D.~M., {Zeebe}, R.~E., \& {Hadden}, S. 2022, \bibinfo{title}{{Stepsize errors in the N-body problem: discerning Mercury's true possible long-term orbits},} \mnras, 510, 4302, \dodoi{10.1093/mnras/stab3664}

\bibitem[{G.~W. Hill(1878)Hill}]{hill1878lunar}
Hill, G.~W. 1878, \bibinfo{title}{Researches in the Lunar Theory,} American Journal of Mathematics, 1, 5, \dodoi{10.2307/2369433}

\bibitem[{A.~R. {Howe} {et~al.}(2026){Howe}, {Becker}, \& {Adams}}]{howe2026host}
{Howe}, A.~R., {Becker}, J.~C., \& {Adams}, F.~C. 2026, \bibinfo{title}{{Architectures of Planetary Systems. II. Trends with Host Star Mass and Metallicity},} \aj, 171, 148, \dodoi{10.3847/1538-3881/ae3aa6}

\bibitem[{J.~D. Hunter(2007)Hunter}]{hunter_2007}
Hunter, J.~D. 2007, \bibinfo{title}{Matplotlib: A 2D graphics environment,} Computing in Science \& Engineering, 9, 90, \dodoi{10.1109/MCSE.2007.55}

\bibitem[{S. {Ida} \& D.~N.~C. {Lin}(2004){Ida} \& {Lin}}]{ida2004deterministic}
{Ida}, S., \& {Lin}, D.~N.~C. 2004, \bibinfo{title}{{Toward a Deterministic Model of Planetary Formation. II. The Formation and Retention of Gas Giant Planets around Stars with a Range of Metallicities},} \apj, 616, 567, \dodoi{10.1086/424830}

\bibitem[{A. {Izidoro} {et~al.}(2021){Izidoro}, {Bitsch}, {Raymond}, {Johansen}, {Morbidelli}, {Lambrechts}, \& {Jacobson}}]{izidoro2021chains}
{Izidoro}, A., {Bitsch}, B., {Raymond}, S.~N., {et~al.} 2021, \bibinfo{title}{{Formation of planetary systems by pebble accretion and migration. Hot super-Earth systems from breaking compact resonant chains},} \aap, 650, A152, \dodoi{10.1051/0004-6361/201935336}

\bibitem[{M. Jette {et~al.}(2002)Jette, Dunlap, Garlick, \& Grondona}]{jette2002slurm}
Jette, M., Dunlap, C., Garlick, J., \& Grondona, M. 2002, \bibinfo{title}{SLURM: Simple Linux Utility for Resource Management,} \url{https://www.osti.gov/biblio/15002962}

\bibitem[{S.~A. {Johnson} {et~al.}(2020){Johnson}, {Penny}, {Gaudi}, {Kerins}, {Rattenbury}, {Robin}, {Calchi Novati}, \& {Henderson}}]{johnson2020ffp}
{Johnson}, S.~A., {Penny}, M., {Gaudi}, B.~S., {et~al.} 2020, \bibinfo{title}{{Predictions of the Nancy Grace Roman Space Telescope Galactic Exoplanet Survey. II. Free-floating Planet Detection Rates},} \aj, 160, 123, \dodoi{10.3847/1538-3881/aba75b}

\bibitem[{M. {Juri{\'c}} \& S. {Tremaine}(2008){Juri{\'c}} \& {Tremaine}}]{juric_tremaine_2008}
{Juri{\'c}}, M., \& {Tremaine}, S. 2008, \bibinfo{title}{{Dynamical Origin of Extrasolar Planet Eccentricity Distribution},} \apj, 686, 603, \dodoi{10.1086/590047}

\bibitem[{P.~C. {Keenan} \& R.~C. {McNeil}(1989){Keenan} \& {McNeil}}]{perkins1989catalog}
{Keenan}, P.~C., \& {McNeil}, R.~C. 1989, \bibinfo{title}{{The Perkins Catalog of Revised MK Types for the Cooler Stars},} \apjs, 71, 245, \dodoi{10.1086/191373}

\bibitem[{D.~R. {Kirsh} {et~al.}(2009){Kirsh}, {Duncan}, {Brasser}, \& {Levison}}]{kirsh2009planetesimal}
{Kirsh}, D.~R., {Duncan}, M., {Brasser}, R., \& {Levison}, H.~F. 2009, \bibinfo{title}{{Simulations of planet migration driven by planetesimal scattering},} \icarus, 199, 197, \dodoi{10.1016/j.icarus.2008.05.028}

\bibitem[{Y. Kozai(1962)Kozai}]{kozai1962secular}
Kozai, Y. 1962, \bibinfo{title}{Secular perturbations of asteroids with high inclination and eccentricity,} \aj, 67, 591

\bibitem[{C. {Lammers} \& J.~N. {Winn}(2024){Lammers} \& {Winn}}]{lammers_winn_2024}
{Lammers}, C., \& {Winn}, J.~N. 2024, \bibinfo{title}{{The Six-planet Resonant Chain of HD 110067},} \apjl, 968, L12, \dodoi{10.3847/2041-8213/ad50d2}

\bibitem[{C. {Lammers} \& J.~N. {Winn}(2026){Lammers} \& {Winn}}]{lammers2025exoplanet}
{Lammers}, C., \& {Winn}, J.~N. 2026, \bibinfo{title}{{On the Exoplanet Yield of Gaia Astrometry},} \aj, 171, 18, \dodoi{10.3847/1538-3881/ae21de}

\bibitem[{J. {Laskar}(1997){Laskar}}]{laskar1997amd}
{Laskar}, J. 1997, \bibinfo{title}{{Large scale chaos and the spacing of the inner planets.},} \aap, 317, L75

\bibitem[{J. {Laskar}(2000){Laskar}}]{laskar2000spacing}
{Laskar}, J. 2000, \bibinfo{title}{{On the Spacing of Planetary Systems},} \prl, 84, 3240, \dodoi{10.1103/PhysRevLett.84.3240}

\bibitem[{E. {Lega} {et~al.}(2021){Lega}, {Nelson}, {Morbidelli}, {Kley}, {B{\'e}thune}, {Crida}, {Kloster}, {M{\'e}heut}, {Rometsch}, \& {Ziampras}}]{lega2021eccentricity}
{Lega}, E., {Nelson}, R.~P., {Morbidelli}, A., {et~al.} 2021, \bibinfo{title}{{Migration of Jupiter-mass planets in low-viscosity discs},} \aap, 646, A166, \dodoi{10.1051/0004-6361/202039520}

\bibitem[{D. {Li} {et~al.}(2020){Li}, {Mustill}, \& {Davies}}]{li2020flyby}
{Li}, D., {Mustill}, A., \& {Davies}, M. 2020, in European Planetary Science Congress, EPSC2020--228, \dodoi{10.5194/epsc2020-228}

\bibitem[{M.~L. Lidov(1962)Lidov}]{lidov1962evolution}
Lidov, M.~L. 1962, \bibinfo{title}{The evolution of orbits of artificial satellites of planets under the action of gravitational perturbations of external bodies,} Planetary and Space Science, 9, 719

\bibitem[{D.~N.~C. {Lin} {et~al.}(1996){Lin}, {Bodenheimer}, \& {Richardson}}]{lin1996orbital}
{Lin}, D.~N.~C., {Bodenheimer}, P., \& {Richardson}, D.~C. 1996, \bibinfo{title}{{Orbital migration of the planetary companion of 51 Pegasi to its present location},} \nat, 380, 606, \dodoi{10.1038/380606a0}

\bibitem[{Y. {Lithwick} \& Y. {Wu}(2014){Lithwick} \& {Wu}}]{lithwick2014secular}
{Lithwick}, Y., \& {Wu}, Y. 2014, \bibinfo{title}{{Secular chaos and its application to Mercury, hot Jupiters, and the organization of planetary systems},} Proceedings of the National Academy of Science, 111, 12610, \dodoi{10.1073/pnas.1308261110}

\bibitem[{T. {Lu} {et~al.}(2025{\natexlab{a}}){Lu}, {An}, {Li}, {Millholland}, {Rice}, {Brandt}, \& {Brandt}}]{lu_hatp11}
{Lu}, T., {An}, Q., {Li}, G., {et~al.} 2025{\natexlab{a}}, \bibinfo{title}{{Planet{\textendash}Planet Scattering and Von Zeipel{\textendash}Lidov{\textendash}Kozai Migration{\textemdash}The Dynamical History of HAT-P-11},} \apj, 979, 218, \dodoi{10.3847/1538-4357/ad9b79}

\bibitem[{T. {Lu} {et~al.}(2024){Lu}, {Hernandez}, \& {Rein}}]{Lu_TRACE}
{Lu}, T., {Hernandez}, D.~M., \& {Rein}, H. 2024, \bibinfo{title}{{TRACE: a code for time-reversible astrophysical close encounters},} \mnras, 533, 3708, \dodoi{10.1093/mnras/stae1982}

\bibitem[{T. {Lu} {et~al.}(2025{\natexlab{b}}){Lu}, {Li}, {Cassese}, \& {Lin}}]{lu_hip}
{Lu}, T., {Li}, G., {Cassese}, B., \& {Lin}, D.~N.~C. 2025{\natexlab{b}}, \bibinfo{title}{{The Dynamical History of HIP-41378 f{\textemdash}Oblique Exorings Masquerading as a Puffy Planet},} \apj, 980, 39, \dodoi{10.3847/1538-4357/ada4b2}

\bibitem[{T. Lu {et~al.}(2023)Lu, Rein, Tamayo, Hadden, Mardling, Millholland, \& Laughlin}]{Lu_2023}
Lu, T., Rein, H., Tamayo, D., {et~al.} 2023, \bibinfo{title}{Self-consistent Spin, Tidal, and Dynamical Equations of Motion in the REBOUNDx Framework,} The Astrophysical Journal, 948, 41, \dodoi{10.3847/1538-4357/acc06d}

\bibitem[{P. Mahalanobis(1936)Mahalanobis}]{mahalanobis1936note}
Mahalanobis, P. 1936, \bibinfo{title}{A note on the statistical and biometric writings of Karl Pearson,} Sankhy{\=a}: The Indian Journal of Statistics (1933-1960), 2, 411

\bibitem[{D. {Malmberg} {et~al.}(2011){Malmberg}, {Davies}, \& {Heggie}}]{malmberg2011flyby}
{Malmberg}, D., {Davies}, M.~B., \& {Heggie}, D.~C. 2011, \bibinfo{title}{{The effects of fly-bys on planetary systems},} \mnras, 411, 859, \dodoi{10.1111/j.1365-2966.2010.17730.x}

\bibitem[{R.~A. {Mardling}(2010){Mardling}}]{mardling_2010}
{Mardling}, R.~A. 2010, \bibinfo{title}{{The determination of planetary structure in tidally relaxed inclined systems},} \mnras, 407, 1048, \dodoi{10.1111/j.1365-2966.2010.16814.x}

\bibitem[{F. {Marzari}(2025){Marzari}}]{marzari2025planet}
{Marzari}, F. 2025, \bibinfo{title}{{Planet-planet scattering in systems of multiple planets of unequal mass},} \mnras, 536, 422, \dodoi{10.1093/mnras/stae2602}

\bibitem[{S. {Millholland} {et~al.}(2017){Millholland}, {Wang}, \& {Laughlin}}]{millholland2017peas}
{Millholland}, S., {Wang}, S., \& {Laughlin}, G. 2017, \bibinfo{title}{{Kepler Multi-planet Systems Exhibit Unexpected Intra-system Uniformity in Mass and Radius},} \apjl, 849, L33, \dodoi{10.3847/2041-8213/aa9714}

\bibitem[{M. Nagasawa \& S. Ida(2011)Nagasawa \& Ida}]{nagasawa2011orbital}
Nagasawa, M., \& Ida, S. 2011, \bibinfo{title}{Orbital distributions of close-in planets and distant planets formed by scattering and dynamical tides,} The Astrophysical Journal, 742, 72

\bibitem[{S. Naoz(2016)Naoz}]{naoz2016eccentric}
Naoz, S. 2016, \bibinfo{title}{The eccentric Kozai-Lidov effect and its applications,} \araa, 54, 441

\bibitem[{D. {Nesvorn{\'y}}(2011){Nesvorn{\'y}}}]{nesvorny2011young}
{Nesvorn{\'y}}, D. 2011, \bibinfo{title}{{Young Solar System's Fifth Giant Planet?},} \apjl, 742, L22, \dodoi{10.1088/2041-8205/742/2/L22}

\bibitem[{M. {Nguyen} \& V. {Adibekyan}(2025){Nguyen} \& {Adibekyan}}]{nguyen2025metallicity}
{Nguyen}, M., \& {Adibekyan}, V. 2025, \bibinfo{title}{{Metallicity Regulates Planet Formation across All Masses},} \aj, 170, 334, \dodoi{10.3847/1538-3881/ae1466}

\bibitem[{A. {Obertas} {et~al.}(2017){Obertas}, {Van Laerhoven}, \& {Tamayo}}]{obertas_2017}
{Obertas}, A., {Van Laerhoven}, C., \& {Tamayo}, D. 2017, \bibinfo{title}{{The stability of tightly-packed, evenly-spaced systems of Earth-mass planets orbiting a Sun-like star},} \icarus, 293, 52, \dodoi{10.1016/j.icarus.2017.04.010}

\bibitem[{T. Oliphant(2006)Oliphant}]{oliphantnumpy}
Oliphant, T. 2006, \bibinfo{title}{{NumPy}: A guide to {NumPy},}, USA: Trelgol Publishing \url{http://www.numpy.org/}

\bibitem[{C. {Petrovich} {et~al.}(2014){Petrovich}, {Tremaine}, \& {Rafikov}}]{petrovich_2014}
{Petrovich}, C., {Tremaine}, S., \& {Rafikov}, R. 2014, \bibinfo{title}{{Scattering Outcomes of Close-in Planets: Constraints on Planet Migration},} \apj, 786, 101, \dodoi{10.1088/0004-637X/786/2/101}

\bibitem[{B. {Pu} \& Y. {Wu}(2015){Pu} \& {Wu}}]{Pu15}
{Pu}, B., \& {Wu}, Y. 2015, \bibinfo{title}{{Spacing of Kepler Planets: Sculpting by Dynamical Instability},} \apj, 807, 44, \dodoi{10.1088/0004-637X/807/1/44}

\bibitem[{S.~N. {Raymond} {et~al.}(2008){Raymond}, {Barnes}, {Armitage}, \& {Gorelick}}]{raymond2008mmrs}
{Raymond}, S.~N., {Barnes}, R., {Armitage}, P.~J., \& {Gorelick}, N. 2008, \bibinfo{title}{{Mean Motion Resonances from Planet-Planet Scattering},} \apjl, 687, L107, \dodoi{10.1086/593301}

\bibitem[{H. {Rein} \& S.~F. {Liu}(2012){Rein} \& {Liu}}]{Rein_2012}
{Rein}, H., \& {Liu}, S.~F. 2012, \bibinfo{title}{{REBOUND: an open-source multi-purpose N-body code for collisional dynamics},} \aap, 537, A128, \dodoi{10.1051/0004-6361/201118085}

\bibitem[{H. {Rein} \& D. {Tamayo}(2015){Rein} \& {Tamayo}}]{rein_2015}
{Rein}, H., \& {Tamayo}, D. 2015, \bibinfo{title}{{WHFAST: a fast and unbiased implementation of a symplectic Wisdom-Holman integrator for long-term gravitational simulations},} \mnras, 452, 376, \dodoi{10.1093/mnras/stv1257}

\bibitem[{H. {Rein} \& D. {Tamayo}(2017){Rein} \& {Tamayo}}]{rein2017simarchive}
{Rein}, H., \& {Tamayo}, D. 2017, \bibinfo{title}{{A new paradigm for reproducing and analyzing N-body simulations of planetary systems},} \mnras, 467, 2377, \dodoi{10.1093/mnras/stx232}

\bibitem[{W.~K.~M. {Rice} {et~al.}(2003){Rice}, {Armitage}, {Bonnell}, {Bate}, {Jeffers}, \& {Vine}}]{rice2003substellar}
{Rice}, W.~K.~M., {Armitage}, P.~J., {Bonnell}, I.~A., {et~al.} 2003, \bibinfo{title}{{Substellar companions and isolated planetary-mass objects from protostellar disc fragmentation},} \mnras, 346, L36, \dodoi{10.1111/j.1365-2966.2003.07317.x}

\bibitem[{G.~R. {Ricker} {et~al.}(2015){Ricker}, {Winn}, {Vanderspek}, {Latham}, {Bakos}, {Bean}, {Berta-Thompson}, {Brown}, {Buchhave}, {Butler}, {Butler}, {Chaplin}, {Charbonneau}, {Christensen-Dalsgaard}, {Clampin}, {Deming}, {Doty}, {De Lee}, {Dressing}, {Dunham}, {Endl}, {Fressin}, {Ge}, {Henning}, {Holman}, {Howard}, {Ida}, {Jenkins}, {Jernigan}, {Johnson}, {Kaltenegger}, {Kawai}, {Kjeldsen}, {Laughlin}, {Levine}, {Lin}, {Lissauer}, {MacQueen}, {Marcy}, {McCullough}, {Morton}, {Narita}, {Paegert}, {Palle}, {Pepe}, {Pepper}, {Quirrenbach}, {Rinehart}, {Sasselov}, {Sato}, {Seager}, {Sozzetti}, {Stassun}, {Sullivan}, {Szentgyorgyi}, {Torres}, {Udry}, \& {Villasenor}}]{ricker2015tess}
{Ricker}, G.~R., {Winn}, J.~N., {Vanderspek}, R., {et~al.} 2015, \bibinfo{title}{{Transiting Exoplanet Survey Satellite (TESS)},} Journal of Astronomical Telescopes, Instruments, and Systems, 1, 014003, \dodoi{10.1117/1.JATIS.1.1.014003}

\bibitem[{V.~S. {Safronov}(1972){Safronov}}]{safronov1972evolution}
{Safronov}, V.~S. 1972, {Evolution of the protoplanetary cloud and formation of the earth and planets.}

\bibitem[{C. {Soubiran} {et~al.}(2010){Soubiran}, {Le Campion}, {Cayrel de Strobel}, \& {Caillo}}]{soubiran2010pastel}
{Soubiran}, C., {Le Campion}, J.~F., {Cayrel de Strobel}, G., \& {Caillo}, A. 2010, \bibinfo{title}{{The PASTEL catalogue of stellar parameters},} \aap, 515, A111, \dodoi{10.1051/0004-6361/201014247}

\bibitem[{G. {Stef{\'a}nsson} {et~al.}(2025){Stef{\'a}nsson}, {Mahadevan}, {Winn}, {Marcussen}, {Kanodia}, {Albrecht}, {Fitzmaurice}, {Mikulskyt{\.{e}}}, {Ca{\~n}as}, {Espinoza-Retamal}, {Zwart}, {Krolikowski}, {Hotnisky}, {Robertson}, {Alvarado-Montes}, {Bender}, {Blake}, {Callingham}, {Cochran}, {Delamer}, {Diddams}, {Dong}, {Fernandes}, {Giovinazzi}, {Halverson}, {Libby-Roberts}, {Logsdon}, {McElwain}, {Ninan}, {Rajagopal}, {Reji}, {Roy}, {Schwab}, \& {Wright}}]{stefansson2025gaia}
{Stef{\'a}nsson}, G., {Mahadevan}, S., {Winn}, J.~N., {et~al.} 2025, \bibinfo{title}{{Gaia-4b and 5b: Radial Velocity Confirmation of Gaia Astrometric Orbital Solutions Reveal a Massive Planet and a Brown Dwarf Orbiting Low-mass Stars},} \aj, 169, 107, \dodoi{10.3847/1538-3881/ada9e1}

\bibitem[{D. {Tamayo} {et~al.}(2020{\natexlab{a}}){Tamayo}, {Rein}, {Shi}, \& {Hernandez}}]{tamayo_reboundx}
{Tamayo}, D., {Rein}, H., {Shi}, P., \& {Hernandez}, D.~M. 2020{\natexlab{a}}, \bibinfo{title}{{REBOUNDx: a library for adding conservative and dissipative forces to otherwise symplectic N-body integrations},} \mnras, 491, 2885, \dodoi{10.1093/mnras/stz2870}

\bibitem[{D. {Tamayo} {et~al.}(2020{\natexlab{b}}){Tamayo}, {Cranmer}, {Hadden}, {Rein}, {Battaglia}, {Obertas}, {Armitage}, {Ho}, {Spergel}, {Gilbertson}, {Hussain}, {Silburt}, {Jontof-Hutter}, \& {Menou}}]{Tamayo20}
{Tamayo}, D., {Cranmer}, M., {Hadden}, S., {et~al.} 2020{\natexlab{b}}, \bibinfo{title}{{Predicting the long-term stability of compact multiplanet systems},} Proceedings of the National Academy of Science, 117, 18194, \dodoi{10.1073/pnas.2001258117}

\bibitem[{D.~P. {Thorngren} {et~al.}(2016){Thorngren}, {Fortney}, {Murray-Clay}, \& {Lopez}}]{thorngren2016mass}
{Thorngren}, D.~P., {Fortney}, J.~J., {Murray-Clay}, R.~A., \& {Lopez}, E.~D. 2016, \bibinfo{title}{{The Mass-Metallicity Relation for Giant Planets},} \apj, 831, 64, \dodoi{10.3847/0004-637X/831/1/64}

\bibitem[{S. {Tremaine}(2023){Tremaine}}]{tremaine_bible}
{Tremaine}, S. 2023, {Dynamics of Planetary Systems}

\bibitem[{D. {Turrini} {et~al.}(2020){Turrini}, {Zinzi}, \& {Belinchon}}]{turrini2020namd}
{Turrini}, D., {Zinzi}, A., \& {Belinchon}, J.~A. 2020, \bibinfo{title}{{Normalized angular momentum deficit: a tool for comparing the violence of the dynamical histories of planetary systems},} \aap, 636, A53, \dodoi{10.1051/0004-6361/201936301}

\bibitem[{S. {van der Walt} {et~al.}(2011){van der Walt}, {Colbert}, \& {Varoquaux}}]{vanderwalt2011numpy}
{van der Walt}, S., {Colbert}, S.~C., \& {Varoquaux}, G. 2011, \bibinfo{title}{{The NumPy Array: A Structure for Efficient Numerical Computation},} Computing in Science and Engineering, 13, 22, \dodoi{10.1109/MCSE.2011.37}

\bibitem[{P. Virtanen {et~al.}(2020)Virtanen, Gommers, Oliphant, Haberland, Reddy, Cournapeau, Burovski, Peterson, Weckesser, Bright, {et~al.}}]{virtanen2020scipy}
Virtanen, P., Gommers, R., Oliphant, T.~E., {et~al.} 2020, \bibinfo{title}{SciPy 1.0: fundamental algorithms for scientific computing in Python,} Nature methods, 17, 261

\bibitem[{H. {von Zeipel}(1910){von Zeipel}}]{von_zeipel_1910}
{von Zeipel}, H. 1910, \bibinfo{title}{{Sur l'application des s{\'e}ries de M. Lindstedt {\`a} l'{\'e}tude du mouvement des com{\`e}tes p{\'e}riodiques},} Astronomische Nachrichten, 183, 345, \dodoi{10.1002/asna.19091832202}

\bibitem[{J.~J. {Wang} {et~al.}(2018){Wang}, {Graham}, {Dawson}, {Fabrycky}, {De Rosa}, {Pueyo}, {Konopacky}, {Macintosh}, {Marois}, {Chiang}, {Ammons}, {Arriaga}, {Bailey}, {Barman}, {Bulger}, {Chilcote}, {Cotten}, {Doyon}, {Duch{\^e}ne}, {Esposito}, {Fitzgerald}, {Follette}, {Gerard}, {Goodsell}, {Greenbaum}, {Hibon}, {Hung}, {Ingraham}, {Kalas}, {Larkin}, {Maire}, {Marchis}, {Marley}, {Metchev}, {Millar-Blanchaer}, {Nielsen}, {Oppenheimer}, {Palmer}, {Patience}, {Perrin}, {Poyneer}, {Rajan}, {Rameau}, {Rantakyr{\"o}}, {Ruffio}, {Savransky}, {Schneider}, {Sivaramakrishnan}, {Song}, {Soummer}, {Thomas}, {Wallace}, {Ward-Duong}, {Wiktorowicz}, \& {Wolff}}]{wang2018hd}
{Wang}, J.~J., {Graham}, J.~R., {Dawson}, R., {et~al.} 2018, \bibinfo{title}{{Dynamical Constraints on the HR 8799 Planets with GPI},} \aj, 156, 192, \dodoi{10.3847/1538-3881/aae150}

\bibitem[{S.~J. {Weidenschilling} \& F. {Marzari}(1996){Weidenschilling} \& {Marzari}}]{weidenschilling1996scattering}
{Weidenschilling}, S.~J., \& {Marzari}, F. 1996, \bibinfo{title}{{Gravitational scattering as a possible origin for giant planets at small stellar distances},} \nat, 384, 619, \dodoi{10.1038/384619a0}

\bibitem[{L.~M. {Weiss} {et~al.}(2018){Weiss}, {Marcy}, {Petigura}, {Fulton}, {Howard}, {Winn}, {Isaacson}, {Morton}, {Hirsch}, {Sinukoff}, {Cumming}, {Hebb}, \& {Cargile}}]{weiss2018peas}
{Weiss}, L.~M., {Marcy}, G.~W., {Petigura}, E.~A., {et~al.} 2018, \bibinfo{title}{{The California-Kepler Survey. V. Peas in a Pod: Planets in a Kepler Multi-planet System Are Similar in Size and Regularly Spaced},} \aj, 155, 48, \dodoi{10.3847/1538-3881/aa9ff6}

\bibitem[{G.~C. {Weldon} {et~al.}(2025){Weldon}, {Naoz}, \& {Hansen}}]{weldon2025cold}
{Weldon}, G.~C., {Naoz}, S., \& {Hansen}, B. M.~S. 2025, \bibinfo{title}{{The Cold Jupiter Eccentricity Distribution is Consistent with EKL Driven by Stellar Companions},} \apjl, 980, L31, \dodoi{10.3847/2041-8213/adb157}

\bibitem[{J. {Wisdom}(2015){Wisdom}}]{wisdom_2015}
{Wisdom}, J. 2015, \bibinfo{title}{{Resolving the Pericenter},} \aj, 150, 127, \dodoi{10.1088/0004-6256/150/4/127}

\bibitem[{R.~A. {Wittenmyer} {et~al.}(2007){Wittenmyer}, {Endl}, \& {Cochran}}]{wittenmyer2007long}
{Wittenmyer}, R.~A., {Endl}, M., \& {Cochran}, W.~D. 2007, \bibinfo{title}{{Long-Period Objects in the Extrasolar Planetary Systems 47 Ursae Majoris and 14 Herculis},} \apj, 654, 625, \dodoi{10.1086/509110}

\bibitem[{R.~A. {Wittenmyer} {et~al.}(2020){Wittenmyer}, {Wang}, {Horner}, {Butler}, {Tinney}, {Carter}, {Wright}, {Jones}, {Bailey}, {O'Toole}, \& {Johns}}]{wittenmyer2020cool}
{Wittenmyer}, R.~A., {Wang}, S., {Horner}, J., {et~al.} 2020, \bibinfo{title}{{Cool Jupiters greatly outnumber their toasty siblings: occurrence rates from the Anglo-Australian Planet Search},} \mnras, 492, 377, \dodoi{10.1093/mnras/stz3436}

\bibitem[{G.-Y. {Xiao} \& F. {Feng}(2025){Xiao} \& {Feng}}]{xiao202514her}
{Xiao}, G.-Y., \& {Feng}, F. 2025, \bibinfo{title}{{Updated Mutual Inclination Measurement for 14 Her b and c},} Research Notes of the American Astronomical Society, 9, 187, \dodoi{10.3847/2515-5172/adef49}

\bibitem[{J.~W. {Xuan} \& M.~C. {Wyatt}(2020){Xuan} \& {Wyatt}}]{XuanWyatt2020}
{Xuan}, J.~W., \& {Wyatt}, M.~C. 2020, \bibinfo{title}{{Evidence for a high mutual inclination between the cold Jupiter and transiting super Earth orbiting {\ensuremath{\pi}} Men},} \mnras, 497, 2096, \dodoi{10.1093/mnras/staa2033}

\bibitem[{D.~A. Yahalomi {et~al.}(2026)Yahalomi, Lu, Armitage, Bedell, Casey, Price-Whelan, \& Rice}]{yahalomi2026astrometric}
Yahalomi, D.~A., Lu, T., Armitage, P.~J., {et~al.} 2026, \bibinfo{title}{The Astrometric Resoeccentric Degeneracy: Eccentric Single Planets Mimic 2: 1 Resonant Planet Pairs in Astrometry,} The Astrophysical Journal Letters, 999, L9

\bibitem[{L. {Yuan} \& M.~H. {Lee}(2024){Yuan} \& {Lee}}]{yuan2024gj1148}
{Yuan}, L., \& {Lee}, M.~H. 2024, \bibinfo{title}{{Scattering of Giant Planets and Implications for the Origin of the Hierarchical and Eccentric Two-planet System GJ 1148},} \apj, 967, 98, \dodoi{10.3847/1538-4357/ad3ba4}

\bibitem[{N.~L. {Zakamska} \& S. {Tremaine}(2004){Zakamska} \& {Tremaine}}]{zakamska2004flyby}
{Zakamska}, N.~L., \& {Tremaine}, S. 2004, \bibinfo{title}{{Excitation and Propagation of Eccentricity Disturbances in Planetary Systems},} \aj, 128, 869, \dodoi{10.1086/422023}

\bibitem[{J. {Zhang} {et~al.}(2025){Zhang}, {Weiss}, {Huber}, {Xuan}, {Bottom}, {Fulton}, {Isaacson}, {MacDougall}, \& {Saunders}}]{ZhangWeiss2025}
{Zhang}, J., {Weiss}, L.~M., {Huber}, D., {et~al.} 2025, \bibinfo{title}{{Discovery of a Jupiter Analog Misaligned to the Inner Planetary System in HD 73344},} \aj, 169, 200, \dodoi{10.3847/1538-3881/ada60a}

\end{thebibliography}
\bibliographystyle{aasjournal}



\end{CJK*}
\end{document}